\documentclass[aip,
 amsmath,amssymb,
 reprint,%
]{revtex4-1}

\usepackage{graphicx}
\usepackage{dcolumn}
\usepackage{bm}

\usepackage[colorlinks]{hyperref}
\usepackage{amsthm}                
\usepackage{algorithm}
\usepackage{caption}
\usepackage{subcaption}
\usepackage{appendix}
\usepackage{booktabs}

\theoremstyle{definition}
\newtheorem{thm}{Theorem}[section]

\newtheorem{alg}[thm]{Algorithm}

\newcommand{\RR}{\mathbb{R}}      

\newcommand{\vecc}{\boldsymbol}

\usepackage[utf8]{inputenc}
\usepackage[T1]{fontenc}
\usepackage{mathptmx}

\begin{document}


\title{Predicting Critical Transitions in Multiscale Dynamical Systems  \\
Using Reservoir Computing}  

\author{Soon Hoe Lim}
\affiliation{Nordita, KTH Royal Institute of Technology and Stockholm University,  106 91 Stockholm, Sweden}

\author{Ludovico Theo Giorgini}
\affiliation{Nordita, KTH Royal Institute of Technology and Stockholm University, 106 91 Stockholm, Sweden}

\author{Woosok Moon}
\affiliation{Nordita, KTH Royal Institute of Technology and Stockholm University, 106 91 Stockholm, Sweden}
\affiliation{Department of Mathematics, Stockholm University, 106 91 Stockholm, Sweden}

\author{J. S. Wettlaufer}
\affiliation{Nordita, KTH Royal Institute of Technology and Stockholm University, 106 91 Stockholm, Sweden}
\affiliation{Yale University, New Haven, Connecticut 06520, USA}

\date{\today}

\begin{abstract}
We study the problem of predicting rare critical transition events for a class of slow-fast nonlinear dynamical systems. The state of the system of interest is described by a slow process, whereas a faster process drives its evolution and induces critical transitions. By taking advantage of recent advances in reservoir computing, we present a data-driven method to predict the future evolution of the state.  We show that our method is capable of predicting a critical transition event at least several numerical time steps in advance. We demonstrate the success as well as the limitations of our method using numerical experiments on three examples of systems, ranging from low dimensional to high dimensional. We discuss the mathematical and broader implications of our results. \\

\end{abstract}

\maketitle

\begin{quotation}
Critical transitions are ubiquitous in nature. These transition events are often induced by a fast driving signal, and are  rare and random. Since such events could lead to significant effects,  it is important to develop effective methods to predict signal-induced critical transitions early. Recently many  studies have been devoted to exploring early warning indicators to predict and characterize the onset of critical transitions. In this paper, we propose and test an alternative method to predict critical transitions within a class of multiscale dynamical systems. Our method is data-driven, inspired by recent advances in reservoir computing, and takes into account the multiscale nature of the systems. We demonstrate the success as well as the limitations of our method using numerical experiments on both low and high dimensional systems. We anticipate that our work will serve as catalyst for further progress in tackling the problem of predicting critical transitions using scientific machine learning.    
\end{quotation}

\section{Introduction}
\label{sect_intro}
The dynamics of many systems in nature are nonlinear, multiscale and noisy, making both the theoretical and numerical modeling and prediction of their states challenging. Of particular interest are those dynamics that often lead to rare transition events. Namely, the system under study spends very long periods of time in various metastable states and only very rarely, and at seemingly random times, does it hop between states. Such sudden changes in the dynamical behavior of complex systems are known as  {\it critical transitions} \cite{scheffer2009early,kuehn2011mathematical}, occurring at so-called {\it tipping points} \cite{ashwin2012tipping}.

One mechanism that explains the hopping behavior is that it is induced by a {\it fast signal} influencing the state of the  otherwise closed system. The usually noisy driving signal comes from an external source. It is often the case that there is a large separation of time scales on which the system state and the signal evolve. Without the driving signal, the system will remain in one state forever. Understanding the dynamics of such system requires us to study the ensemble of transition paths between the different metastable states \cite{weinan2006towards,forgoston2018primer,hartmann2014characterization}.

The above signal-induced phenomenon can be modeled by a class of non-autonomous open dynamical systems, whose state $x: [0,T] \to \RR^n$, $T>0$, is described by: 
\begin{align} 
\dot{x}(t) &= F(x(t),t) + f(t), \label{gen}  
\end{align}
where $F(x(t),t)$ is a force field, providing a deterministic backbone, and $f(t)$ is a fast driving noisy signal, describing small perturbations imparted to the system. 
In principle, one does not have  a priori knowledge of the mathematical model for $f(t)$, but only has access to data for $x$ and, possibly, a mathematical model for $F(x(t),t)$ at their disposal.  The unknown $f(t)$ is generally a random function, and it could be regular, chaotic or stochastic. It is challenging to infer an accurate model for $f(t)$ using only the data for $x$. In fact, in many cases one could construct several models that are capable of  describing the data equally well, and often there are ambiguities in the choice of model. An example that illustrates this issue is the problem of distinguishing deterministic chaos from stochasticity \cite{kaplan1993coarse,agarwal2016maximal}. Indeed, on one hand, when certain assumptions are satisfied, chaotic systems can be well approximated by a stochastic one \cite{beck1990brownian,mackey2006deterministic,chevyrev2016multiscale}. On the other hand, many stochastic systems can be described by a chaotic model \cite{gaspard1998experimental}. In the absence of a uniquely defined model for $f(t)$, one has to resort to a  model-free, data-driven approach, which is currently an extremely active area in applied dynamical systems. We refer the reader to the book by Brunton and Kutz \cite{brunton_kutz_2019} for an excellent review of data-driven methods.

An important and practical question concerning the system \eqref{gen} is the following -- can we predict the future evolution of the system state using only its data? Moreover, can we achieve this for a sufficiently long time, and to a desired level of accuracy and certainty? Note that this is different from asking one to infer a mathematical model (which may have a very low predictive power, as for example in the case of a random walk model) from the data. Typically, the data is complex and multiscale in nature, therereby complicating analysis and prediction. {\it The main goal of this work is to propose and test a machine learning based solution to predict rare events in multiscale noisy nonlinear dynamical systems, making use of only  the (slow) system state data and a partial knowledge of the physics of the generating system.} Since the occurrence of rare events can have significant deleterious or positive implications, it is important to quantify and predict them in advance to inform decision making \cite{giorgini2020precursors} -- this is the main motivation of this paper. To the best of our knowledge this is the first attempt to solve this particular problem using a machine learning method. Here we will focus on the case where the system of interest is one-dimensional. While we focus on prediction of rare events, our method can be applied to predicting future states in general multiscale systems.


By exploiting recent advances in machine learning, we construct an algorithm to solve the prediction problem. The field of machine learning is experiencing a major recent resurgence of interest, with wide ranging applications and significant implications in many areas of science and engineering \cite{mehta2019high,carleo2019machine}. We note in particular neural networks and deep learning, whose industry wide applications have been made possible due to availability of large amount of data and advances in computer hardware development \cite{goodfellow2016deep}. For instance, by training on a sufficiently large set of data, one can classify handwritten digits to unprecedented accuracy \cite{lecun1998gradient}, predict and analyze time series \cite{vlachas2018data}, infer the H\"older exponent of stochastic processes \cite{stone2020calibrating}, characterize anomalous diffusion \cite{Bo19}, learn how to construct linear embeddings of nonlinear dynamics  \cite{lusch2018deep}, replicate chaotic attractors and calculate Lyapunov exponents \cite{pathak2017using}, solve high-dimensional nonlinear PDEs \cite{raissi2017physics}, and others that are too numerous to list here. As powerful as they appear, we emphasize that it is by no means an easy task to apply the tools of machine learning to real world data sets and one needs to proceed with caution and avoid potential pitfalls\cite{riley19}. Indeed, one of the main challenges in machine learning deals with the ability of the algorithm to be generalized to unseen data. Moreover, the empirical approach of machine learning is refined only with practices that have principally been discovered by trial and error. Our approach falls within the scope of scientific machine learning \cite{baker2019workshop}, an emerging research area that is focused on the opportunities and challenges of machine learning in the context of applications across science and engineering.


This paper is organized as follows. In Section \ref{sect_problem}, we motivate and describe the class of dynamical systems of interest from which the data is generated. They are special cases of \eqref{gen} with a specified model for $f(t)$. We then introduce the problem of data-based rare event prediction. In Section \ref{sect_alg}, we  present and explain in detail a reservoir computing based method to solve the prediction problem using a variant of the echo state network.   In Section \ref{sect_exp}, we apply and test the method to predict rare events in three different systems. We make concluding remarks in Section \ref{sect_concl}. The Appendix contains  further details on training, validation (model selection) and testing.   


\section{Multiscale Noisy Systems and the Prediction Problem} \label{sect_problem}

We consider scenarios where the available data is generated by the following family of continuous-time slow-fast systems (parametrized by $\epsilon > 0$) \cite{Pavliotis,rodenbeck2001dynamical,berglund2006noise}: 
\begin{align} 
\dot{x}(t) &= F(x(t),t) + \frac{\sigma}{\epsilon} G(\xi(t)), \label{slow} \\
\dot{\xi}(t) &= \frac{1}{\epsilon^2} H(\xi(t)), \label{fast}
\end{align}
with the initial conditions $x(0)=x_0$ and $\xi(0)=\xi_0$, where $x_0$ and $\xi_0$ can be either fixed or random. In the case when both initial conditions are fixed, \eqref{slow}-\eqref{fast} describe a deterministic dynamical system; otherwise, it is a random dynamical system.  In the above, $x: I \to \RR$ is a slow process, $\xi: I \to \RR^m$ is a fast process (noise),  $F: \RR \times I \to \RR$ is a deterministic force field, $G: \RR^m \to \RR$ is a functional (observable) of the fast process, $H: \RR^m  \to \RR^m$ is a deterministic vector field describing the fast process, $\sigma>0$ is a small constant controlling the strength of influence of the fast process on the slow one, and $I $ is a time interval on which the system evolves. Here $F, G$ and $H$ are $O(1)$ as $\epsilon \to 0$. In many applications of interest, $F$, $G$ and $H$ may be highly nonlinear. The assumption that the data is modeled by the above systems is not too restrictive, for as is very often the case in the analysis of experimental data, there is not a unique single model for the system generating the data.

Often one is interested in the dynamics of the slow process and the case where the driving signal, $f(t) = \frac{\sigma}{\epsilon} G(\xi(t))$, is a stochastic process such as Gaussian white noise.  A crucial property of  dynamical systems of the form \eqref{slow}-\eqref{fast} is that  stochastic behavior emerges as $\epsilon$ becomes smaller. Indeed, it can be rigorously shown that under appropriate assumptions on the initial conditions, $F$,  $G$  and $H$,  $x(t)$ converges in law (homogenizes) to $X(t)$ as $\epsilon \to 0$ for $t \in I$, where $X(t)$ is a diffusion process solving the stochastic differential equation (SDE) \cite{melbourne2011note,chevyrev2016multiscale}:
\begin{equation}
dX(t) = F(X(t),t) dt + \tilde{\sigma} dW(t).
\end{equation}
Here $\tilde{\sigma}>0$ is a constant (involving the time integral of the correlation of the fast process) and $W(t)$ is a Wiener process. Usually a mixing assumption is imposed on the fast flow.  However, such an assumption is not necessary for homogenization. The system only needs to be sufficiently chaotic. See, for instance, Remark 2.1 in \cite{gottwald2013homogenization} and also \cite{kelly2016smooth,kelly2017deterministic,chevyrev2016multiscale}.

Therefore, the family of equations $\eqref{slow}$-$\eqref{fast}$ can be viewed as approximations of a stochastic model.  Such a perspective has been adopted to study a variety of noisy systems \cite{arnold2001hasselmann,berner2017stochastic,mitchell2012data,hasselmann1976stochastic}.  We refer to Section \ref{sect_exp} for concrete examples. An important class of $\xi(t)$ are those that exhibit deterministic chaos, which has been studied and observed \cite{gaspard1998experimental} in a range of physical systems. Note that $G$ can be generalized to be dependent on $x(t)$, in which case the limiting SDE will have a multiplicative noise, but we will not pursue this case here.

We now formulate the prediction problem introduced in Section \ref{sect_intro} systematically. First, we lay out our assumptions. The only data available to us is that for $x$, which is assumed to be generated by the systems \eqref{slow}-\eqref{fast}, and we do not have access to the data for $\xi$.  Let $I=[t_{init},t_{obs}] \cup [t_{obs}, t_f]$, where $t_{init}  < t_{obs} < t_f$. Here $t_{init}$ denotes an initial time, $t_{obs}$ denotes an observation time, beyond which we do not have access to system state, and $t_f$ denotes the final time at which the prediction will be made. Suppose that we are given a sufficiently long and high-frequency time series  for $x$ on $[t_{init}, t_{obs}]$. Furthermore, we are blind to the actual mathematical model for the fast process. However, a partial knowledge of the physics of the system of interest is known. In particular, it is assumed that we know the exact expression describing the force field $F$. This assumption is satisfied  when one can reconstruct the force field from the data accurately, which may be possible in many practical situations \cite{garcia2018high}.

We are given a time series data for $x$,  a sequence $(x_k)_{k=0,\dots,N} =  (x(t_0), x(t_1), x(t_2), \dots, x(t_N))$, where $t_0 = t_{init}$, $t_i = t_{init} + i \Delta t $ ($i=1, \dots, N$) are the sampling times, $\Delta t$ is the time step size, $t_N = t_{init} + N \Delta t =: t_{obs}$, and $N+1$ is the number of available samples. Our time series does not record the occurrence of a rare event and we assume that a rare event will occur shortly after time $t_{obs}$. Any precursors for this rare event must then be hidden in the  time series.  We then attempt to answer the following questions. 
\begin{itemize}
\item[(1)] Can we predict if and when a rare event will occur in a given future time window?  Can we infer the characteristics of the event?
\item[(2)] How far in advance can we predict the rare event?
\item[(3)] With what accuracy and certainty can we achieve these goals?
\item[(4)] Is it possible to answer all of these questions with a computationally inexpensive method and/or using a relatively short time series data for $x$?
\end{itemize}
Clearly these are challenging questions. The degree of difficulty depends on the characteristics of the dynamical systems generating the data for $x$. For a given amount of data the difficulty increases  as $\epsilon$ becomes smaller, in which case the statistical behavior of the driving noise is closer to that of a white noise and so predictability is lost in the limit. Thus, analysis of the data should be performed on a case by case basis.

\section{A Machine Learning Based Prediction Method} \label{sect_alg}

\subsection{The method} \label{subsect_method}
We first present a three-step procedure that will allow us to investigate the questions posed at the end of Section \ref{sect_problem}. This procedure lies at the heart of our method. We then discuss a number of heuristic issues associated with our approach. 

\begin{alg} {\it Predicting rare critical transition events.} \label{thealg}

\noindent Under the assumptions, setting and notation described in Section \ref{sect_problem}:
\begin{itemize}
\item[(S1)] {\bf Feature extraction.} Extract the fast driving signal using the data for $x$ and the known expression for $F$:
\begin{equation}
f(t_i) = \frac{x(t_{i+1})-x(t_i)}{\Delta t} - F(x(t_i),t_i), 
\end{equation}
for $i = 0,1, \dots, N-1$, where $\Delta t$ is a uniform time step.
\item[(S2)] {\bf Machine learning. } Using $(f(t_i))_{i=0,\dots,N-1}$ as the training data, predict the values of $f$ for $M+1$ time steps into the future (i.e., beyond time $t_{obs}-\Delta t$) using a supervised learning algorithm (for instance, the Algorithm \ref{alg_desn} in Section \ref{sect_exp}) that is best fit for the task to infer  $(f(t_{N}), f(t_{N+1}), \dots, f(t_{N+M}))$.  
\item[(S3)] {\bf Numerical integration.} Numerically evolve the system \eqref{slow} up to time $t_{N+M}$, with $(f(t))_{t=t_0, \dots, t_{N+M}}$ in place of $\frac{\sigma}{\epsilon} G(\xi(t))$ in \eqref{slow}, using the step size $\Delta t$. The predicted values for $x$ in the time window $[t_{obs}+\Delta t, t_f = (N+M)\Delta t]$ are then obtained from the resulting numerical solutions.
\end{itemize}
\end{alg}

Algorithm \ref{thealg} is the method that we propose and use for predicting rare events. It allows the prediction of the system state $M$ time steps beyond the observation time $t_{obs}$, the result of which can be used to investigate the questions posed in Section \ref{sect_problem}. To be able to answer these questions with a desired level of confidence, the predicted values should be as close as possible to the target (actual) values, i.e. the generalization (out-of-sample) error should be small. 

Before we discuss the details of implementation for each of the three steps above, a few remarks are in order. A natural approach is to  apply a suitable machine learning algorithm directly to the data for $x$ and attempt to predict the future states. While this seems like a sensible approach, it is unrealistic to expect an algorithm to learn the multiscale nature of the data accurately. In fact, it is not clear beforehand which algorithm is best and in many cases the rare event will not be detected successfully  (see Section \ref{sect_exp}). Our method circumvents this challenge and provides an alternative route for handling multiscale data. Moreover, the simplicity of our method provides an additional advantage.

The quality and accuracy of the prediction results will rely strongly on how well each step in the algorithm is executed. Errors will accumulate as one progresses through these steps. Indeed, (S1) involves a numerical approximation of the driving signal.  In (S2), errors will arise from both the use of the training data (where the numerical errors from (S1) are hidden) as well as the machine learning algorithm itself. In (S3), an additional error due to numerical integration is inevitable.  Provided that the accumulation of these errors is negligibly small and well controlled, one can learn and predict the system states with reasonably good accuracy, as demonstrated with the examples in Section \ref{sect_exp}. Rigorous error analysis is not the focus of the present paper  and so will  not be presented here.

Steps (S1) and (S3) are straightforward to implement, so we must discuss (S2),  whose implementation is the most challenging part of the method. We will formulate this step as the problem of learning the training data $(f(t_i))_{i=0,\dots,N-1}$ with parametrized high-dimensional nonlinear dynamical systems. 




\subsection{Echo state network (ESN) and its deep version} \label{sect_esn}

There are many machine learning algorithms that one can use  to implement step (S2). Algorithms from deep learning include convolutional neural networks, recurrent neural networks (RNNs), and encoder-decoder networks, each of which can be implemented using  various architectures and training schemes \cite{goodfellow2016deep}. As nonlinear state space models, RNNs have {\it dynamical memory}, which means that they are capable of preserving in their internal state a nonlinear transformation of the input history. They are, therefore, particularly well suited to deal with sequential data. We will implement (S2) using a type of brain-inspired RNN known as the {\it echo state network} (ESN).  For practical introductions and technical details on ESNs, we refer the reader to Refs. \cite{lukovsevivcius2012practical,jaeger2002tutorial}.

ESN belongs to the paradigm of reservoir computing \cite{tanaka2019recent} and is computationally less costly to train than other variants of RNNs, which typically use a backpropagation through time algorithm for gradient descent based training  \cite{jaeger2004harnessing,lukovsevivcius2009reservoir}.
 Similar to other RNNs, the ESN can, under fairly mild and general assumptions, be shown to be a universal approximator of arbitrary dynamical systems \cite{grigoryeva2018echo}. In contrast to standard design and training schemes for RNNs, but conceptually similar to the kernel methods (c.f. \cite{lim2020understanding,tino2020dynamical}), the neural network (called the reservoir) in ESNs is generated randomly and only the readout from the reservoir is trained. The outputs are linear combinations of the internal states and possibly the inputs and a constant (bias). This reduces the training to solving a linear regression problem, minimizing the mean squared error between the outputs and the target values.  
 
It is crucial that typically the number of reservoir elements in an ESN is larger than that used in RNNs trained with a gradient descent based method, resulting in an over-parametrized neural network. The key observation is that with sufficient over-parametrization, gradient descent based methods will implicitly leave some weights describing the network relatively unchanged, so the optimization dynamics will behave as if those weights are essentially fixed at their initial values \cite{NIPS2019}. Fixing these weights explicitly leads to the approach of learning with random reservoir features. Therefore, we can successfully learn with gradient descent based trained neural networks whenever we can successfully learn with ESNs.

Even though not all the weights of the network are trained, it has been shown that ESNs work surprisingly well and achieve excellent performance in many benchmark tasks, including winning the NN3 financial time series prediction challenge \cite{ilies2007stepping}.  Recently ESNs  have been shown to predict chaotic systems remarkably well \cite{pathak2017using,Pathak18_spatiotemp,pathak2018hybrid,zimmermann2018observing,lu2018attractor}. They may outperform other machine learning algorithms in certain prediction tasks. For instance, it has been shown that the ESN substantially outperforms the deep feed-forward neural network and the RNN with long short-term memory (LSTM) for predicting short-term evolution of a multiscale spatio-temporal Lorenz-96 system \cite{npg-27-373-2020} (see also \cite{vlachas2020backpropagation}).  Moreover, ESNs do not suffer from the vanishing and exploding gradient problem typically encountered when training other RNNs \cite{jaeger2002tutorial}.  These results motivate our choice of (a variant of the) ESN over other machine learning methods. We emphasize, however, that the ESN may not be the most optimal network for our prediction task and we remain mindful of its  shortcomings, in particular  its sensitive dependence on the hyperparameters. We leave careful comparison of different machine learning methods for our prediction problem to future work.

To achieve our goals, we also use a deep version of the ESN (see also the discussion in  Section \ref{sect_exp}(b)) whenever needed. Our deep echo state network (DESN) consists of  organized hierarchically stacked ESNs, whose architecture and training algorithm will be described in the following subsubsections. Such deep ESNs are more expressive than shallow ESNs, in the sense that they are able to develop in their internal states a multiple time-scale representation of the temporal information \cite{gallicchio2017deep,gallicchio2020deep}. We remark that in contrast to feed-forward neural networks, it is often not obvious how one should construct a deep RNN \cite{pascanu2013construct}. In particular, different variants of deep echo state networks can be constructed, depending on the task at hand.

\subsubsection{Architecture of the ESN and its deep version}
Similar to other RNNs, the ESN is a parametrized, high-dimensional, discrete-time, non-autonomous, nonlinear state-space model, describing a dynamical input-output relation:
\begin{align}
\vecc{x}(t_{n+1}) &= \vecc{f}(\vecc{x}(t_n), \vecc{u}(t_{n}), \vecc{y}(t_n), \vecc{\nu}(t_n), \vecc{\theta}_{\vecc{f}} ),  \\
\vecc{y}(t_{n+1}) &= \vecc{g}(\vecc{x}(t_{n+1}), \vecc{u}(t_{n+1}), \vecc{\nu}(t_{n+1}), \vecc{\theta}_{\vecc{g}} ), 
\end{align}
for $n=0,1,\dots, N-1$. Here  $\vecc{u}(t) \in \RR^{n_u}$ is the input  at time $t$, $\vecc{x}(t) \in \RR^{n_x}$ is the internal/hidden state of the reservoir, $\vecc{y}(t) \in \RR^{n_y}$ is the output, $\vecc{\nu}(t) \in \RR^{n_\nu}$ is an external perturbation (noise/regularization),  $\vecc{f}$ and $\vecc{g}$ are generally nonlinear functions, and $\vecc{\theta}_{\vecc{f}}$ and $\vecc{\theta}_{\vecc{g}}$ are model parameters. The network non-linearly embeds the input into a higher dimensional space where
the original problem is more likely to be solved linearly.

Our ESN is a specific implementation of the above state-space model, putting constraints on a fully connected RNN and {\it with the inputs $(\vecc{u}(t_0), \dots, \vecc{u}(t_{N-1}))$ set to be a training time series} during the training phase.  In the following, $t_n = n \Delta t$, for $n=0,\dots,N-1$,  where $\Delta t$ is a fixed time step size. 
If the number of layers, $n_L:=L+1$, is chosen to be one, i.e., a shallow ESN, then we have the following update equations for the states and outputs:
\begin{align}
\vecc{x}^{(0)}(t_{n+1}) &= \tanh(\vecc{W}^{(0)}\vecc{x}^{(0)}(t_n)  + b \vecc{W}_{b}  + \vecc{W}_{in} \vecc{u}(t_n)) + \nu \vecc{\xi}_n^{(0)}, \label{run1} \\ 
\vecc{y}(t_{n+1}) &= \vecc{W}_{out} (\vecc{x}^{(0)}(t_{n+1}); b), \label{up1}
\end{align}
for $n = 0, 1, \dots, N-1$,  with  $\vecc{x}(t_0) := \vecc{x}^{(0)}(t_0) = \vecc{0}$. In the above, we have used the following vector concatenation notation: $(\vecc{a}; \vecc{b}) := (a_1, a_2, \dots, a_n, b_1, b_2, \dots, b_m) \in \RR^{n+m}$, for two vectors $\vecc{a}=(a_1, a_2, \dots, a_n) \in \RR^n$ and $\vecc{b} = (b_1, b_2, \dots, b_m)\in \RR^m$. Also, we have used the following convention for component-wise application of the activation function: $\tanh(\vecc{a}):=(\tanh(a_1), \dots, \tanh(a_n))\in \RR^n$ for $\vecc{a}=(a_1,\dots,a_n)\in \RR^n$.

Otherwise, we employ the following deep version of ESN, which we refer to as DESN:
\begin{widetext}
\begin{align}
\vecc{x}^{(0)}(t_{n+1}) &= \tanh(\vecc{W}^{(0)}\vecc{x}^{(0)}(t_n) + b \vecc{W}_{b})  + \nu \vecc{\xi}^{(0)}_n, \label{run2} \\
\vecc{x}^{(l)}(t_{n+1}) &= \tanh(\vecc{W}^{(l)}\vecc{x}^{(l)}(t_n) + \vecc{V}^{(l)} \vecc{x}^{(l-1)}(t_{n})) + \nu \vecc{\xi}^{(l)}_n, \ \ \text{ for } l=1,\dots,L-1,  \\
\vecc{x}^{(L)}(t_{n+1}) &= \tanh(\vecc{W}^{(L)}\vecc{x}^{(L)}(t_n) + \vecc{V}^{(L)}\vecc{x}^{(L-1)}(t_{n}) + \vecc{W}_{in} \vecc{u}(t_n)) + \nu \vecc{\xi}^{(L)}_n,  \label{run3} \\   
\vecc{y}(t_{n+1}) &= \vecc{W}_{out}(\vecc{x}^{(0)}(t_{n+1}); \dots; \vecc{x}^{(L)}(t_{n+1}); b), \label{up2}
\end{align}
\end{widetext} 
\noindent for $n = 0, 1, \dots, N-1$,  with the initial condition $\vecc{x}(t_0) :=  (\vecc{x}^{(0)}(t_0); \dots; \vecc{x}^{(L)}(t_0)) = \vecc{0}$.

In \eqref{run1}-\eqref{up2}, the training input $\vecc{u}$ and output $\vecc{y}$ come from a compact subset of $\RR^{n_u}$ and of $\RR^{n_y}$ respectively, the vector $\vecc{x}^{(i)} \in \RR^{n_{x_i}}$ (for $i=0,\dots,L$) is the $i$th hidden state, the constant $b \in \RR$ introduces bias, the matrices $\vecc{W}^{(i)} \in \RR^{n_{x_i} \times n_{x_i}}$ (for $i=0,\dots,L$) , $\vecc{W}_{b} \in \RR^{n_{x_0}}$, $\vecc{V}^{(i)} \in \RR^{n_{x_{i}} \times n_{x_{i-1}}}$  (for $i=1,\dots,L$) and  $\vecc{W}_{in} \in \RR^{n_{x_{L}} \times n_u}$ are fixed internal connection weights whose entry values are set to random values, the matrix $\vecc{W}_{out} \in \RR^{n_y \times  (n_{x_0} + \dots + n_{x_L} + 1)}$ is the readout weight matrix whose entries are to be learned. The vectors $\vecc{\xi}^{(k)}_n$, $k=0,\dots,L$, $n=0,1,\dots,N-1$, are i.i.d. random vectors describing added noise during the sampling at each layer and  $\nu$ is a noise (regularization) intensity parameter (we have taken the same noise level for each layer).  The activation function in each layer is taken to be a (vectorized) hyperbolic tangent function. At each training step, the input is fed into the first layer, which is then connected to the next layer via the connection weights. The ansatz for the output at each update time is taken to be a linear combination of the elements of the hidden states and a bias value.

\subsubsection{Algorithm for training the ESN and its deep version}

The above implementation gives a randomly constructed RNN prior to training. It may generally develop oscillatory or chaotic behavior even in the absence of external excitation by the input, and therefore the subsequent network states, starting from an arbitrary state $\vecc{x}(t_0)$, may not converge to the zero state. To ensure that the ESN/DESN converges to the desired state, the internal connection weights are scaled such that the resulting (untrained) input-driven recurrent network (or the ``dynamical reservoir'') is appropriately stabilized or ``damped'',  forcing it to have the so-called {\it ``echo state property"}. The echo state property ensures  that the current network state is  uniquely determined by the history of the input  provided that the RNN has been run for a sufficiently long time (see \cite{jaeger2002tutorial,manjunath2013echo} and the references therein for details, subtleties, and other equivalent conditions for the echo state property). Once this initialization is made, we can proceed to the training phase and then the prediction phase, to be described in the following.

\begin{figure*}
\caption{{\it A schematic of the ESN/DESN and the workflow during the training phase described in Section \ref{sect_esn} (Algorithm \ref{thealg}).} In our numerical experiments, the training time series $(u(t_0), \dots, u(t_{N-1}))$ is the fast driving signal extracted from $(x(t_0), \dots, x(t_{N}))$. }
\centering
\includegraphics[scale=0.8]{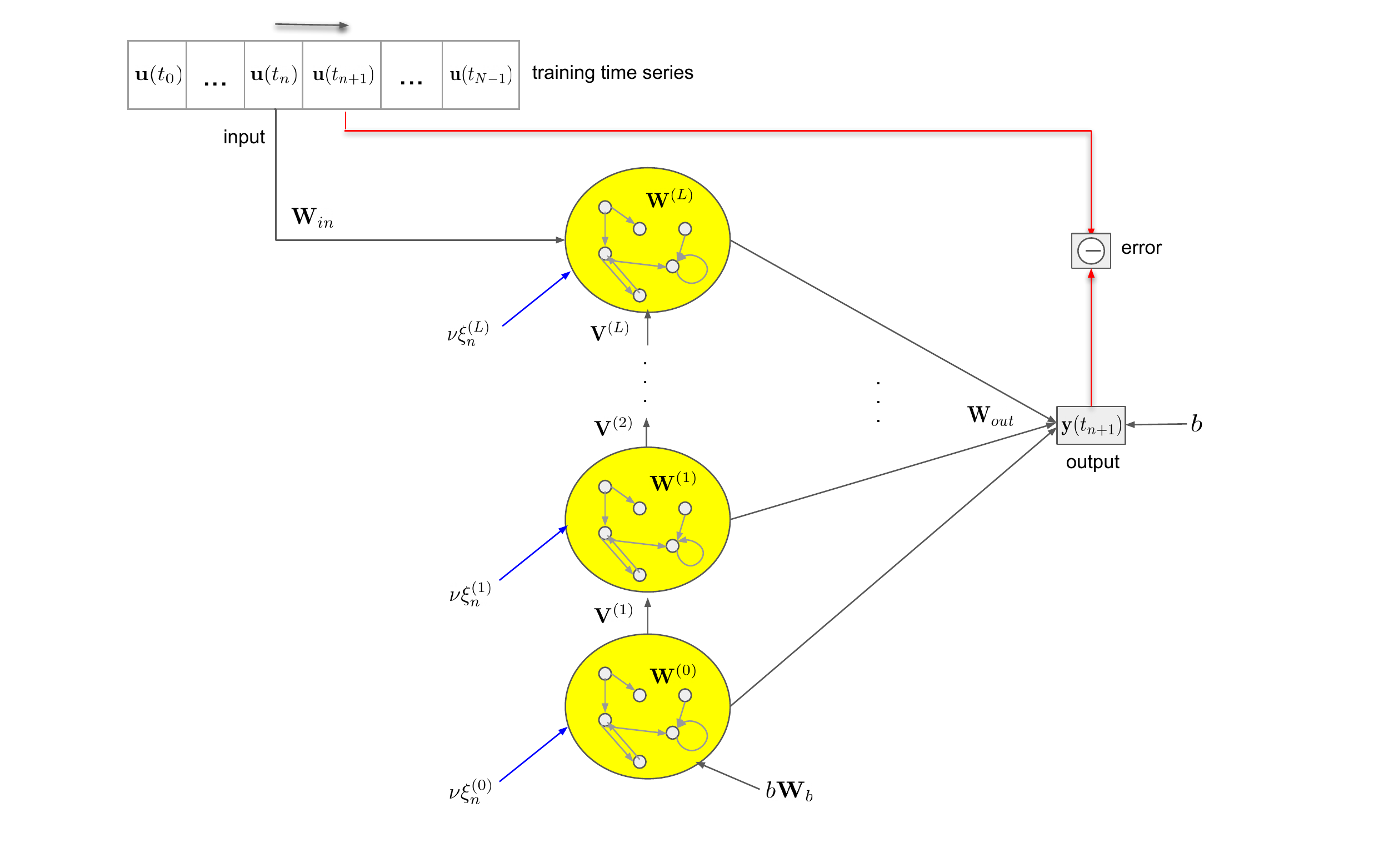}
\label{nn_pic}
\end{figure*}

We now give a complete description of setting up, training and using the ESN and its deep version for the prediction task. It is based closely on the techniques developed in \cite{jaeger2002tutorial}.  For a schematic of a general ESN/DESN architecture and the training process, we refer to Figure \ref{nn_pic}.

\begin{alg} {\it Initializing, training, and using the ESN ($L=0$) and DESN ($L \geq 1$).} \label{alg_desn}

\noindent Given a training set consisting of the input sequence $(\vecc{u}(t_0), \dots, \vecc{u}(t_{N-1}))$, find a trained ESN/DESN parametrized by $(\vecc{W}^{(i)}, \vecc{W}_b,  \vecc{V}^{(j)}, \vecc{W}_{in}, \vecc{W}_{out}, \nu)_{i,j =0 \dots,L}$ whose network output $(\vecc{y}(t_0), \dots, \vecc{y}(t_{N-1}))$ approximates the  input sequence. 
\begin{itemize}
\item[(1)] {\bf  Initialize the ESN/DESN  to ``ensure''\footnote{Doing so will usually, but not always, ensure that the resulting network satisfies the echo state property \cite{jaeger2002tutorial}.} the echo state property is satisfied.}
\item[(1a)] Depending on the training data (length, difficulty of task, etc.), select appropriate dimensions/sizes (i.e., $n_{x_0},\dots,n_{x_L}$) for the connection weight matrices. These dimensions are hyperparameters that can be tuned. The matrix elements of these matrices are then selected randomly as follows:
\begin{align}
(\vecc{W}_{b})_{kl} &\sim \mathrm{Unif}(-1,1), \\
(\vecc{V}^{(j)})_{kl} &\sim \mathrm{Unif}(-0.5,0.5) \text{  , for }  j=1,\dots,L, \\
(\vecc{W}^{(i)})_{kl} &\sim \mathrm{Unif}(-0.5,0.5) \text{ ,   for }  i=0,\dots,L, \\
(\vecc{W}_{in})_{kl} &\sim \mathrm{Unif}(-1,1).
\end{align}
\item[(1b)] These matrices are rescaled as follows:
\begin{widetext}
\begin{equation}
\vecc{W}_{b}, \vecc{V}^{(j)}, \vecc{W}^{(i)}, \vecc{W}_{in} \mapsto  \frac{\vecc{W}_{b}}{\| \vecc{W}_{b} \|+0.001},  \frac{\vecc{V}^{(j)}}{\|\vecc{V}^{(j)}\| + 0.001}, \frac{\vecc{W}^{(i)}}{\|\vecc{W}^{(i)}\|+0.001 }, \frac{\vecc{W}_{in}}{\| \vecc{W}_{in} \|+0.001},
\end{equation}
\end{widetext}
\noindent where $\| \cdot \|$ denotes the Frobenius norm. 
\item[(1c)] Set a fraction of  elements (connection weights) in the matrices $\vecc{W}^{(i)}$ to zero (the fraction, denoted $r_i$, chosen is a hyperparameter for sparsity of the matrices used for each layer) and then rescale the resulting matrices $\tilde{\vecc{W}}^{(i)}$ appropriately using their spectral radius, i.e.,
\begin{equation}
\tilde{\vecc{W}}^{(i)} \mapsto  \rho^{(i)}_{des} \frac{\tilde{\vecc{W}}^{(i)}}{\rho^{(i)}},  
\end{equation}
where $\rho^{(i)}$ is the spectral radius of $\tilde{\vecc{W}}^{(i)}$, $\rho^{(i)}_{des}$  is the desired spectral radius, and $i=0,\dots,L$. The desired spectral radius chosen is  less than one to ensure contractivity of the dynamics and is another tunable hyperparameter. The sparsity hyperparameter is chosen in such a way that  sufficiently rich  dynamics of different internal units/hidden states can be obtained.

\noindent (1a)-(1c)  then give an untrained network (dynamical reservoir) that satisfies (typically in practice) the required properties. We  initialize the network state with $\vecc{x}(t_0) = \vecc{0}$. \\

\item[(2)] {\bf Train the readout by solving a least squares linear regression problem. }
\item[(2a)] If needed, discard an initial transient by disregarding the first $i_{transient} = \min(\mathrm{int}(N/10),100)$ states, where $\mathrm{int}(x)$ is the integer part of $x$ and $\min(x,y)$ is the minimum of $x$ and $y$. 
\item[(2b)] Run the network on the entire input (training) sequence and collect the output sequence, i.e. according to \eqref{run1}-\eqref{up1} for the shallow ESN and \eqref{run2}-\eqref{up2} for the DESN. During sampling we have added a small amount of noise or regularization, whose intensity is determined by the hyperparameter $\nu$, to stabilize the network and prevent overfitting \cite{lukovsevivcius2012practical}. We choose the elements of the i.i.d. noise vectors $\vecc{\xi}_n^{(k)}$ to be distributed according to  $\mathrm{Unif}(-0.5,0.5)$.  
\item[(2c)] Solve the least squares regression problem:
\begin{equation} \label{lsp}
\min_{\vecc{W}_{out}} \frac{1}{N-i_{transient}} \sum_{i=i_{transient}}^{N-1} \|\vecc{W}_{out} \tilde{\vecc{x}}(t_i) - \vecc{u}(t_i)\|^2,
\end{equation}
where $\tilde{\vecc{x}}(t_i)=(\vecc{x}(t_i); b) \in \RR^{n_{x_0} + \dots + n_{x_L} + 1}$. We remind the reader that $\vecc{W}_{out} \in \RR^{n_y \times (n_{x_0}+\dots + n_{x_L}+1)}$ is the only trainable matrix in the ESN/DESN.  As the above problem admits a closed form solution, the solution, denoted $\vecc{W}^{opt}_{out}$, can be obtained directly by applying the Moore-Penrose pseudo inversion as follows.

Let $\tilde{\vecc{X}} =  [\vecc{X}  \ \ \vecc{b}] \in \RR^{(N-i_{transient})  \times (n_{x_0}+\dots+n_{x_L}+1)}$, with $\vecc{X} \in \RR^{ (N-i_{transient}) \times  (n_{x_0}+\dots + n_{x_L})}$ the design matrix, i.e.  a block matrix stacked vertically with the matrices

\noindent  $[\vecc{x}^{(j)}(t_{i_{transient}}) \ \  \hdots \ \  \vecc{x}^{(j)}(t_{N-1}) ]^T  \in \RR^{ (N-i_{transient}) \times  n_{x_j}} $ ($j=0, \dots, L$), $\vecc{b} = (b, b, \dots, b) \in \RR^{N-i_{transient}}$, and $\vecc{U} = [\vecc{u}(t_{i_{transient}}) \ \ \hdots \ \  \vecc{u}(t_{N-1})]^T  \in \RR^{(N-i_{transient})  \times n_y}$. Then the solution to \eqref{lsp} can be obtained as: 
\begin{equation}
\vecc{W}^{opt}_{out} = (\tilde{\vecc{X}}^{+} \vecc{U} )^T, 
\end{equation}
where $+$ and $T$  denote the Moore-Penrose inverse and transposition respectively.  This completes the training phase. \\

\item[(3)] {\bf Run the trained ESN/DESN autonomously for prediction.} 

\noindent  During the prediction phase, we use $\vecc{y}(t_m) = \vecc{W}_{out}^{opt}(\vecc{x}(t_m); b)$ in place of $\vecc{u}(t_m)$ for $m \geq N-1$ in the update equation \eqref{run3} for DESN (or \eqref{run1} for ESN). We then propagate the  trained network forward in time according to the resulting update equation. This allows us to obtain the predicted values $(\vecc{y}(t_{N}), \dots, \vecc{y}(t_{N+M}))$, where $\vecc{y}(t_i)  = \vecc{W}_{out}^{opt} \tilde{\vecc{x}}(t_i)$, for $i=N, \dots, N+M$. 
\end{itemize}
\end{alg}

The smaller the generalization error  on the prediction time horizon, the more accurate is the prediction.  The quality of prediction achieved by our ESN/DESN  depends on the selection of hyperparameters, which need to be chosen following an appropriate model selection procedure, which adapts to the data set on hand. The tunable hyperparameters are $n_L := L+1$ (number of layers in the DESN), the $n_{x_i}$ (dimension of the connection weight matrices at layer $i+1$), the $r_i$ (sparsity parameter), the $\rho^{(i)}_{des}$ (the desired spectral radius of these matrices at layer $i+1$), and $\nu$ (noise intensity). We remark that the above algorithm gives a specific way to initialize and train the ESN/DESN. Other variants may also be considered \cite{lukovsevivcius2009reservoir}.  To implement (S2) in Algorithm \ref{thealg}, we apply Algorithm \ref{alg_desn} to the training data $(f(t_i))_{i=0,\dots,N-1}$. Together with (S1) and (S3), this completes the description of our rare event prediction method.

Lastly, we discuss a physically motivated intuition about ESNs (as can be said for DESNs). On one hand, ESNs can be constructed by sampling from a class of continuous-time dynamical systems \cite{sherstinsky2020fundamentals}, which satisfy a universal property in the sense that they can approximate any continuous-time system on compact time intervals to arbitrary degree of precision. On the other hand, the data itself are generated from a nonlinear chaotic system. The ESN approach amounts to learning the data by feeding the data as input (signal) into an ESN, thereby inducing an interaction between the two dynamical systems. In our case, we would like the output of the network to reproduce the input time series, and good  performance could be obtained under suitable conditions (when the hyperparameters are tuned optimally).  From a physical point of view, these conditions cause the two dynamical systems to synchronize, and so learning here corresponds to finding optimal conditions to achieve generalized synchronization of the system generating the training data and a signal-fed ESN \cite{PhysRevE.99.042203,pecora2015synchronization}. Indeed, the idea of synchronization between two dynamical systems has  been exploited for time series prediction \cite{PhysRevLett.101.154102}. It remains interesting to understand how out-of-sample performance of ESNs depend on the characteristics of input data and the noise (either those present in the data or those added to regularize the training) using the notion of generalized synchronization.




\section{Numerical Experiments}
\label{sect_exp}

We apply the method presented in Section \ref{sect_alg} to study the questions  posed in Section \ref{sect_problem} for three data sets generated by dynamical systems of different complexity.  

From the data set in each example, a training set (with $N+1-N_v$  data points), a validation set (with $N_v$ data points) and a test set (with $M$ data points) are assigned before we apply the method. We are going to perform an ensemble based prediction by utilizing multiple independently  trained  networks.  We take weighted averaged values of the predicted values produced by these networks as the final predicted values. We refer to Appendix, in particular Figure  \ref{ms} there, for details of  training and prediction procedure.  We emphasize that while a comprehensive investigation of dependence of the results obtained here on the choice of hyperparameters is interesting on its own, this is not our focus here.

Throughout this section, all variables considered are real and one-dimensional.

\subsection{Example 1: A bi-stable system, driven by a fast Lorenz-63 system} \label{eg1}

\noindent {\bf Data generation.} The data for $x$ is generated by the following slow-fast system \cite{givon2004extracting}:
\begin{align}
\dot{x}(t) &= x(t)[1-x^2(t))]  + \frac{\sigma}{\epsilon} y_2(t), \label{x1} \\
\dot{y}_1(t) &= \frac{10}{\epsilon^2}[y_2(t)-y_1(t)], \label{lor1} \\
\dot{y}_2(t) &= \frac{1}{\epsilon^2}[28y_1(t)-y_2(t)-y_1(t) y_3(t))], \label{lor2} \\
\dot{y}_3(t) &= \frac{1}{\epsilon^2}[y_1(t) y_2(t) - \frac{8}{3} y_3(t))]. \label{lor3}
\end{align}

In \eqref{x1}-\eqref{lor3}, $x$ is the state of the system of interest  and its evolution is driven by a fast chaotic signal $y_2/\epsilon$, which is modeled as follows. The vector state $(y_1, y_2, y_3)$ is described by the Lorenz-63 model  with the classical parameter values that lead to chaotic behavior \cite{lorenz1963deterministic}. At these parameter values, $y_2$ is ergodic with invariant measure supported on a set of zero volume. The equation for $x$ is therefore an ODE driven by a fast chaotic signal with characteristic time $\epsilon^2$. 

\begin{figure}[!h]
\caption{{\it Time series data for $x$ in Example 1 up to time $t=100$}, of which only a segment prior to time $t=37 < \tau_0 = 38.52$ (i.e., at most $N+1=3701$ data points) will be available for training. Note that the sample path of $x$ first crosses zero at $t = \tau_0$ (indicated by the yellow dashed line), where a critical transition occurs.}
\centering
\includegraphics[width=0.48\textwidth]{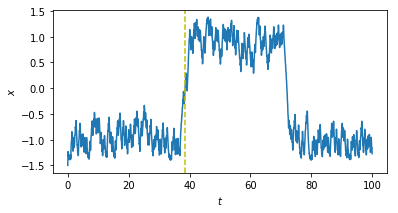}
\label{eg1_whole}
\end{figure}

To generate the data, we use a uniform time step of $\Delta t= 0.01$ to integrate \eqref{x1}-\eqref{lor3} with $\sigma = 0.08$, $\epsilon =0.5$, $x(0)=-1.5$, and $y_i(0) \sim \mathrm{Unif}(-10,10)$ for $i=1,2,3$, up to time $t=100$.  The autocorrelation time (defined as the time $\tau$ at which the autocorrelation function $R(\tau)=1/e$) of samples of the driving signal is estimated to be $0.05$.  Note that the time step is chosen to be small enough so that we can sample the scale on which the fast driving signal takes place. The time series generated for $x$ is plotted in Figure \ref{eg1_whole}. \\

We remark that repeating the above data generation using a  different numerical time step, initial conditions (or random seeds) and parameters will produce time series with different characteristics. For instance, the resulting time series will display a different number of critical transitions occurring at different times. We assume that we only have access to a segment of the {\it single} time series plotted in Figure \ref{eg1_whole} up to time $t<\tau_0$. \\

\noindent {\bf Generating system: dynamics and applications.}
Applying the discussion in Section \ref{sect_problem}, the family of systems \eqref{x1}-\eqref{lor3} (parametrized by $\epsilon$) can be viewed as an approximation to the Markovian system:
\begin{equation}
dX(t) = X(t)[1-X^2(t))] dt + \tilde{\sigma}dW(t),
\end{equation}
where $\tilde{\sigma}$ is an effective diffusion constant and $W(t)$ is a Wiener process, in the sense that as $\epsilon$ becomes smaller $x(t)$ converges in law to the process $X(t)$ solving the SDE above. Recall that we have chosen $\epsilon = 0.5$ for data generation and therefore the data can be thought of coming from an approximately stochastic system.

In the absence of the driving signal, the equation for $x$ has two stable fixed points, at $x=-1$ and $x=1$, and an unstable one at $x=0$. Starting from an initial state, the system will eventually evolve towards a nearby stable state.  The presence of noise alters this dynamics, causing an occasional transition of the system between stable states. In the case where the noise amplitude is small, such a transition is a rare event, occurring at a seemingly unpredictable time (see Figure \ref{eg1_whole}). In our case, $x$ starts near the fixed point $x=-1$ and will eventually jump to that at $x=1$, the prediction of which is of great interest. From the data, we find that the first crossing time of the sample path to zero is $\tau_0 := \inf\{t \in [0,100] : x(t) > 0\} = 38.52$. Our goal is then to predict an approaching rare event using only (a segment of) the data consisting of time series up to time $t <  \tau_0$.

From a statistical mechanical point of view, the system \eqref{x1}-\eqref{lor3} describes an overdamped Brownian particle  moving in a symmetric double-well potential. In this case, $x$ is the position of the particle and time-integrated $y_2$ models the fluctuations due to its interaction with the environment. The above model for $x$ is  also often used in climate physics, an example of which is to view $x$ as the sea-surface temperature anomaly and $\xi=\frac{\sigma}{\epsilon} y_2$ as the impact of small-scale atmospheric variability \cite{hasselmann1976stochastic,mitchell2012data,berner2017stochastic}. \\


\noindent {\bf Results and discussion.}  The prediction results for different training scenarios using Algorithm \ref{thealg} are displayed in the figures in Table \ref{res_eg1}-\ref{std_eg1}. For each case where a fixed number of training data points is used, we present two prediction models, each of which depends on the size $(N_v)$ of the validation set  used (see Appendix for details).

\begin{table*}
 \caption{{\it Prediction results using the data set in Example 1 under various scenarios.} \\ 
$N$ is the number of data points in the training set, $N_v$ is the number of data points in the validation set (see Appendix).  In each figure: the target trajectory for $x$ (in red), $N_{eff}$ predicted trajectories from $N_{eff}$ independently trained and selected models, and the averaged predicted trajectory (in thick blue) up to several time steps into the future. The shaded light blue region represents the  $90\%$ confidence interval of the predicted values.}
        \centering
        \begin{tabular}{|c|c|c|}
        \hline
           \toprule
            Scenarios & $N_v = 10$ & $N_v = 20$  \\
            \hline
            \midrule
            Baseline  & (0a) \includegraphics[width=0.4\textwidth]{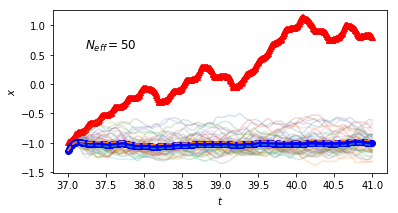} & (0b) \includegraphics[width=0.4\textwidth]{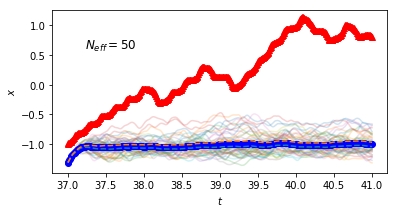} \\ \hline
            $N=3700$ & (1a) \includegraphics[width=0.4\textwidth]{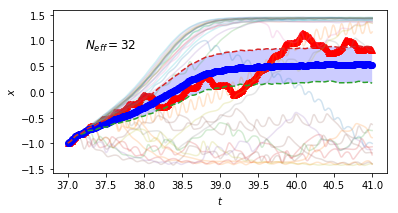} & (1b) \includegraphics[width=0.4\textwidth]{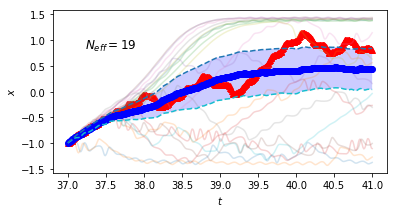} \\ \hline
            $N=3690$ & (2a) \includegraphics[width=0.4\textwidth]{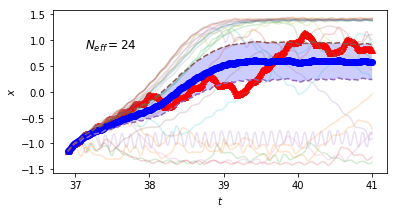} & (2b) \includegraphics[width=0.4\textwidth]{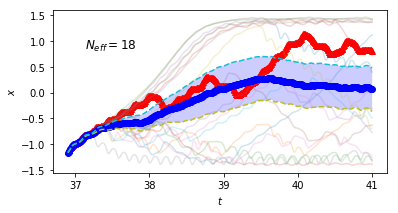} \\ \hline
            $N=3680$ & (3a) \includegraphics[width=0.4\textwidth]{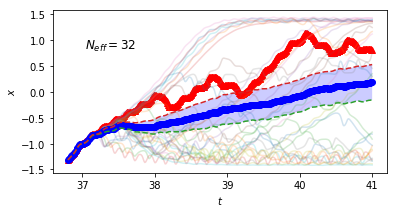} & (3b) \includegraphics[width=0.4\textwidth]{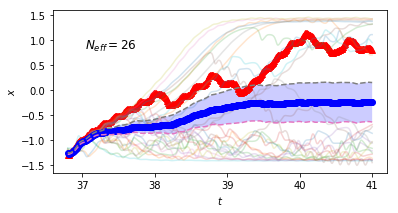} \\ \hline
            $N=3670$ & (4a) \includegraphics[width=0.4\textwidth]{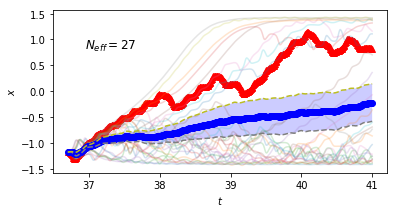} & (4b) \includegraphics[width=0.4\textwidth]{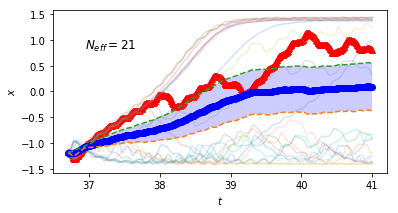}  \\
            \bottomrule
            \hline
        \end{tabular}
        \label{res_eg1}
    \end{table*}

\begin{table*}
 \caption{{\it   Error of the predicted results and its mean value (in thick blue) for Example 1. }}
         \centering
        \begin{tabular}{|c|c|c|}
        \hline
           \toprule
            Scenarios & $N_v = 10$ & $N_v = 20$  \\
            \hline
            \midrule
              $N=3700$ & (1a) \includegraphics[width=0.4\textwidth]{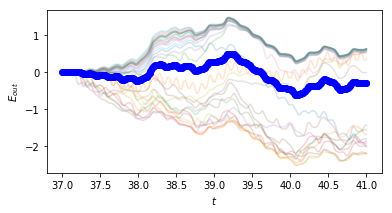} &(1b)  \includegraphics[width=0.4\textwidth]{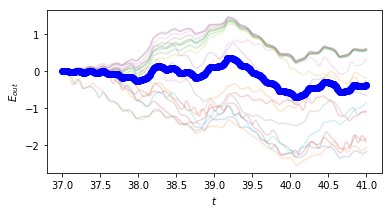} \\ \hline
            $N=3690$ & (2a) \includegraphics[width=0.4\textwidth]{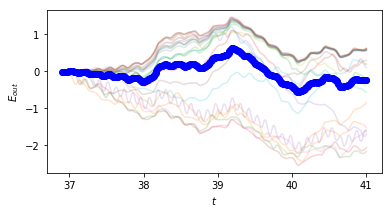} & (2b) \includegraphics[width=0.4\textwidth]{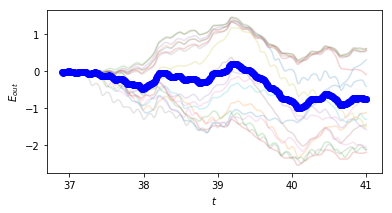} \\ \hline
            $N=3680$ & (3a) \includegraphics[width=0.4\textwidth]{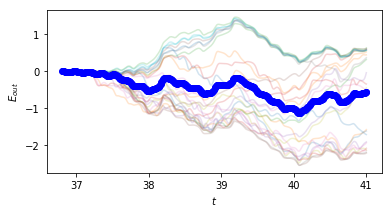} & (3b) \includegraphics[width=0.4\textwidth]{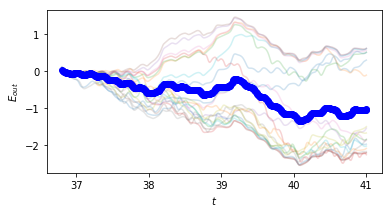} \\ \hline
            $N=3670$ & (4a) \includegraphics[width=0.4\textwidth]{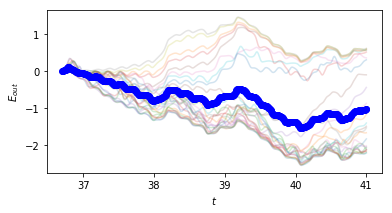} & (4b) \includegraphics[width=0.4\textwidth]{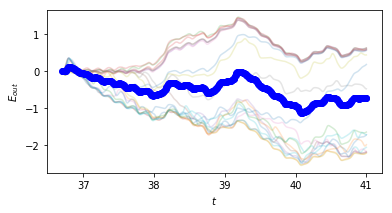}  \\
            \bottomrule
            \hline
        \end{tabular}
        \label{error_eg1}
    \end{table*}

\begin{table*}
 \caption{{\it  Standard deviation of the errors for Example 1. } 
(a) $N=3700$; (b) $N=3690$; (c) $N=3680$; (d) $N=3670$. }
         \centering
        \begin{tabular}{|c|c|}
        \hline
         (a)  \includegraphics[width=0.4\textwidth]{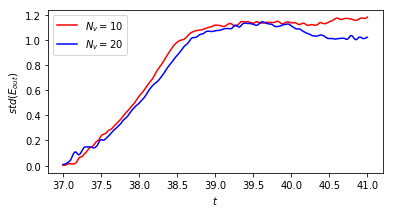}  &
         (b) \includegraphics[width=0.4\textwidth]{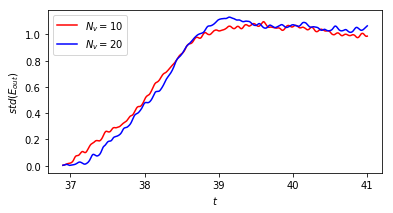}  \\ \hline
         (c) \includegraphics[width=0.4\textwidth]{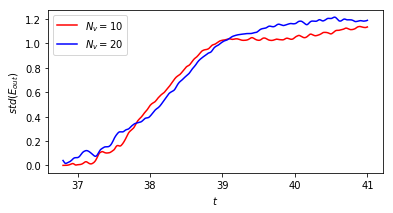}  
  &     (d) \includegraphics[width=0.4\textwidth]{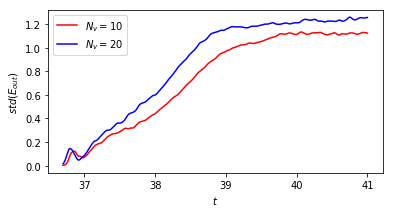} \\
            \hline
        \end{tabular}
        \label{std_eg1}
    \end{table*}

Figure (1a)-(1b) and Figure (2a)-(2b) in Table \ref{res_eg1} show that our method is capable of quite confidently predicting the rare event and its transition path at least 162 numerical time steps in advance, i.e., prior to first crossing of the slow process to zero at $\tau_0 = 38.52$. Or equivalently, at least about 32  autocorrelation time of the fast driving signal.   In particular, the target future trajectory lies principally  within the $90$ percent confidence interval of the predicted trajectory.  Given that a relatively short time series was used for training, it would appear that the accuracy of these predictions is remarkable. 


Figure (3a)-(4b) show how the inability of the ESN to make accurate long-term predictions makes accurate prediction of the {\it entire} transition path (i.e., up to time $t=40$) from an earlier time more difficult. Indeed, the longer into the future the prediction is, the less confident the results are. This is shown by the growth of the standard deviation of the prediction errors, which seems to saturate about a maximum value that increases with fewer training data points.  However, the results in Figure (3a) and Figure (4b) are still respectable, as they indicate that the trajectory will cross the origin during the prediction time window, albeit not able to trace the target transition path precisely. 

These figures also depict the difference in the prediction results obtained with different values of $N_v$ used (notably in the figures (2a)-(4b)), showing  intricacies of the machine learning based method. However, the qualitative behavior of mean value of the error of the predicted results are essentially the same for different $N_v$ (see the figures in Table \ref{error_eg1}). The same can be said for the standard deviation of these errors, which grows rapidly on the prediction horizon where the rare transition occurs and then approaches a  plateau after that (see the figures in Table \ref{std_eg1}). This plateau is slightly higher in all the cases where $N_v = 20$ than those where $N_v=10$, indicating the higher variance of the prediction models which set aside more data points for validation. However, for the case of $N=3670$, the model with the higher variance give better predictive performance than that with the lower variance, and is  able to predict a cross-over to the origin during the prediction interval.

Figure (0a)-(0b) confirm  that our method is far better than the direct method of applying an ESN to the data for $x$, which we take as a baseline result for comparison. We emphasize that one can also apply other machine learning algorithms such as the gradient descent based RNN and  convolutional neural network to implement the direct method \cite{bengio2009learning}. However, after some experiments we found that the predictive performance is similar to the baseline result, i.e., they fail to predict the approaching rare event, rendering the prediction task almost impossible using the direct method. This comparison study enhances the veracity of our method, which exploits the crucial idea of appropriately taking into account the multiscale nature of the system generating the data. The success of our method in predicting the rare transition event lies in its ability to separate the slow and fast components of the data with the help of some physical knowledge about the generating system.

We remark that if we run the trained networks  much further into the future,  the predicted trajectory, not suprisingly, fails to capture the second transition that occurs around $t=71$. However, we will be able to predict the second transition confidently if more data points are used for training. All the results discussed so far are obtained by training on a trajectory that starts from the initial time ($t_{init}=0$) and ends at a time before a rare event occurs. A natural question is whether one can achieve prediction results of comparable quality using a shorter trajectory that comes sufficiently close to the rare event but starts at a later time $t_{init}>0$. Given the chaoticity of the generating system with a known predictability horizon determined by the Lyapunov exponent, one expects the answer to this question is affirmative and can indeed be supported by numerical experiments.

Lastly, we emphasize that our problem is more challenging than that of predicting the second component of a Lorenz-63 system from numerical data generated by the system itself. Even though we expect the extracted driving signal here is representative of  second component of the Lorenz-63 system (up to a scaling factor) and so one would expect that good prediction results are achievable (in lights of recent results \cite{ pathak2017using}), there are inherent errors  in the extraction process (S1) and so here we are really dealing with a noise corrupted version of the data. In this case, the prediction could be highly non-trivial (and perhaps impossible if the errors are large enough) due to possible dominance of noise in certain segments of the reconstructed time series. 




\subsection{Example 2: A tri-stable system, with periodic forcing and a fast Ornstein-Uhlenbeck-like process}

\begin{figure}[!h]
\caption{{\it Time series data for $x$ in Example 2 up to time $t=100$}, of which only a segment prior to $t=94.1 < \tau_1 = 96.16$ (i.e., at most $N+1=9411$ data points) will be available for training. Note that the sample path of $x$ first crosses one at $t = \tau_1$ (indicated by the yellow dashed line), where a critical transition occurs.}
\centering
\includegraphics[width=0.48\textwidth]{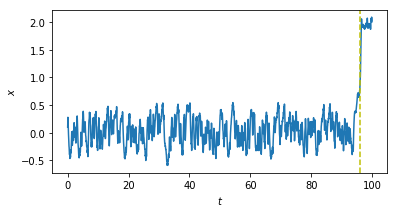}
\label{eg2_whole}
\end{figure}

\noindent {\bf Data generation.} The data for $x$ is generated by the following slow-fast system:
\begin{align}
\dot{x}(t) &= x(t)[1-x(t)][1+x(t)][x(t)-2][x(t)+2] \nonumber \\
&\ \ \ \ \ \  +   A \cos(\omega t) + \sigma_0 z(t), \label{eg2_1} \\
\dot{z}(t) &= -\alpha_1 z(t) + \frac{\sigma_1}{\epsilon} y_2(t), \label{eg2_2}
\end{align}
where $x$ describes the state of the system of interest whose evolution is driven by a fast Ornstein-Uhlenbeck like signal $z$, and $y_2$ is the second component of the Lorenz-63 system \eqref{lor1}-\eqref{lor3}. For data generation,  we use a uniform time step of $\Delta t= 0.01$ to integrate \eqref{eg2_1}-\eqref{eg2_2}  with $x(0)=0.1$, $z(0), y_i(0) \sim \mathrm{Unif}(-10,10)$ (for $i=1,2,3$), $A= 0.5$, $\omega = 2\pi$, $\epsilon=0.5$, $\sigma_0 = 0.2$, $\alpha_1=1000$, and $\sigma_1=1000 \epsilon$, up to time $t=100$. The autocorrelation time  of samples of the driving signal is estimated to be $0.07$. The time step is small enough to sample the scale on which the fast signal takes place. 
The time series generated for $x$ is plotted in Figure \ref{eg2_whole}. We assume that we only have access to a segment of the {\it single} time series plotted in Figure \ref{eg2_whole} up to time $t<\tau_1$.
\\

\noindent {\bf Generating system: dynamics and applications.} The system \eqref{eg2_1}-\eqref{eg2_2}, is more complex than that in Example 1, of which it is an extended version, in the sense that the force field is generalized to include a time-dependent external force and the driving signal is described by a higher dimensional system. One can view  \eqref{eg2_1}-\eqref{eg2_2} as a family of systems (parametrized by $\epsilon$) approximating the following SDE system:
\begin{align}
dX(t) &= X(t)[1-X(t)][1+X(t)][X(t)-2][X(t)+2] dt \nonumber \\ 
&\ \ \ \ \ \ \ + A\cos(2 \pi t)dt + \sigma_0 Z(t) dt, \label{aoup1} \\
dZ(t) &= - \alpha_1 Z(t) dt + \tilde{\sigma}_1 dW(t), \label{aoup2}
\end{align}
where $\tilde{\sigma}_1$ is an effective diffusion constant and $W(t)$ is a Wiener process. As $\epsilon$ (chosen to be $0.5$ for data generation here) becomes smaller, $(x(t),z(t))$ converges in law to $(X(t),Z(t))$ solving the non-autonomous SDE system \eqref{aoup1}-\eqref{aoup2}.

In contrast to  Example 1, in the absence of the driving signal $z(t)$, equation \eqref{eg2_1} for $x$ has three stable periodic orbits centered at $x=-2,0,2$ and two unstable ones centered at $x=-1,1$. Here our system state starts in the middle potential well and will, due to influence of the driving signal as well as the periodic forcing, transits to one of the left or right nearby stable orbits at a random time (see Figure \ref{eg2_whole}). It is thus natural to ask which potential well will the system transit to and at what time will the transition occur. From the data, we find that the first crossing time of the sample path to one is $\tau_1 := \inf\{t \in [0,100] : x(t) > 1\} = 96.16$. 

The addition of the periodic forcing $A\cos(2\pi t)$ introduces another time scale into the system. When this time scale is of the same order of the mean exit time from the potential (the Kramers time), a resonance-like mechanism where the noise can lead to the amplification of the periodic signal takes place. This resonance is induced by a chaotic signal, and is closely related to stochastic resonance, a noise-induced phenomenon first introduced in the context of climate modeling \cite{benzi1981mechanism,benzi1982stochastic}, which has been found to occur in many physical and biological systems \cite{gammaitoni1998stochastic}.

One example of such systems describes the dynamics of an overdamped self-propelled active particle, which converts energy absorbed from the environment into a directed motion, rendering the system out of equilibrium. The position of the particle can be described by $X$ in \eqref{aoup1} and the active force or self-propulsion is modeled by $Z$ in \eqref{aoup2}, in which case the particle is trapped in a triple-well potential and subject to periodic forcing. This is a variant of the model of an active Ornstein-Uhlenbeck process  widely used to study active matter \cite{romanczuk2012active}. \\

\noindent {\bf Results and discussion.} The prediction results for different training scenarios using Algorithm \ref{thealg} are displayed in Figures \ref{res_eg2}-\ref{std_eg2}. For each case where a fixed number of training data points is used, we present two prediction models, each of which depends on the size $(N_v)$  of the  validation set used (see Appendix for details).

\begin{table*}
 \caption{{\it Prediction results using the data set in Example 2 under various scenarios.} \\ 
$N$ is the number of data points in the training set, $N_v$ is the number of data points in the validation set (see Appendix).  
In each figure: the target trajectory for $x$ (in red), $N_{eff}$ predicted trajectories from $N_{eff}$ independently trained and selected models, and the averaged predicted trajectory (in thick blue) up to several time steps into the future. The shaded light blue region represents the  $90\%$ confidence interval of the predicted values.}
        \centering
        \begin{tabular}{|c|c|c|}
        \hline
           \toprule
            Scenarios & $N_v = 4$ & $N_v = 8$  \\
            \hline
            \midrule
            Baseline  & (0a) \includegraphics[width=0.4\textwidth]{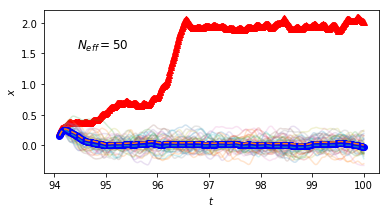} & (0b) \includegraphics[width=0.4\textwidth]{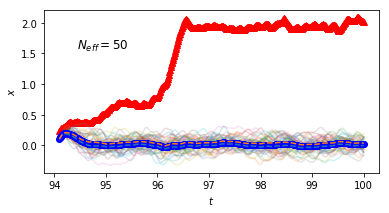} \\ \hline
            $N=9410$ & (1a) \includegraphics[width=0.4\textwidth]{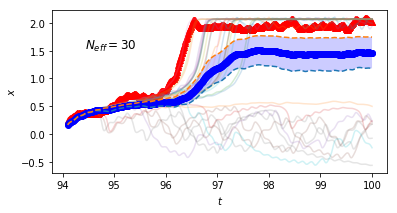} & (1b) \includegraphics[width=0.4\textwidth]{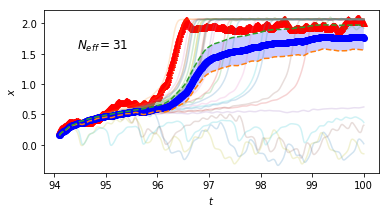} \\ \hline
            $N=9400$ &  (2a) \includegraphics[width=0.4\textwidth]{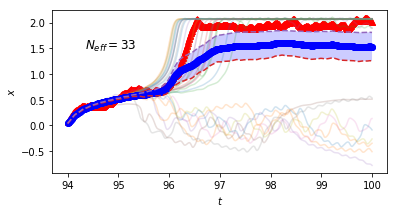} & (2b) \includegraphics[width=0.4\textwidth]{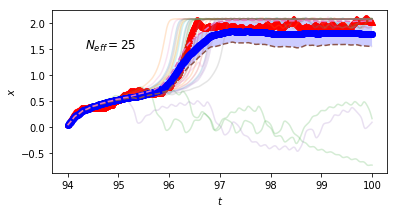} \\ \hline
            $N=9390$ &  (3a) \includegraphics[width=0.4\textwidth]{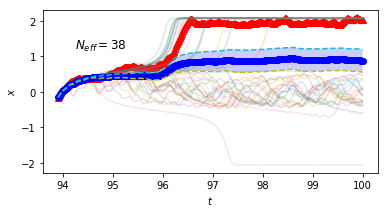} & (3b) \includegraphics[width=0.4\textwidth]{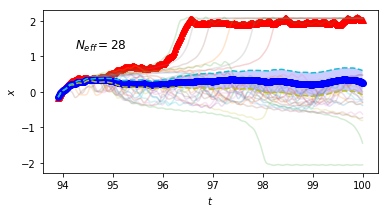} \\
            \bottomrule
            \hline
        \end{tabular}
        \label{res_eg2}
    \end{table*}

\begin{table*}
 \caption{{\it   Error of the predicted results and its mean value (in thick blue) for Example 2. }}
         \centering
        \begin{tabular}{|c|c|c|}
        \hline
           \toprule
            Scenarios & $N_v = 4$ & $N_v = 8$  \\
            \hline
            \midrule
            $N=9410$ & (1a) \includegraphics[width=0.4\textwidth]{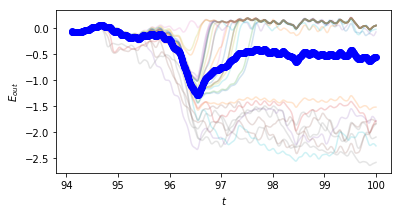} & (1b) \includegraphics[width=0.4\textwidth]{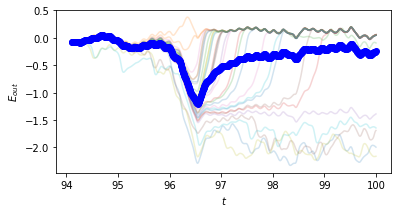} \\ \hline
            $N=9400$ & (2a) \includegraphics[width=0.4\textwidth]{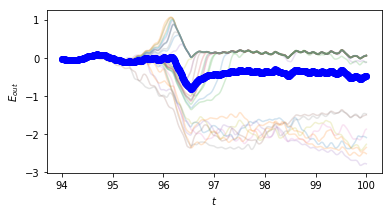} & (2b) \includegraphics[width=0.4\textwidth]{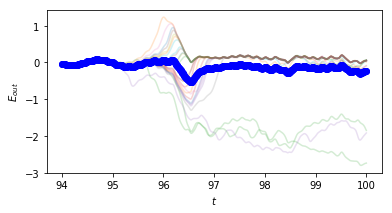} \\ \hline
            $N=9390$ & (3a) \includegraphics[width=0.4\textwidth]{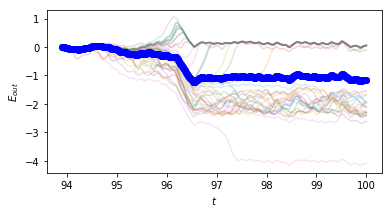} & (3b) \includegraphics[width=0.4\textwidth]{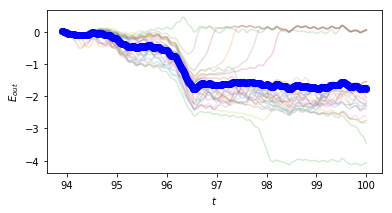} \\
            \bottomrule
            \hline
        \end{tabular}
        \label{error_eg2}
    \end{table*}

\begin{table*}
 \caption{{\it  Standard deviation of the errors for Example 2. } 
(a) $N=9410$; (b) $N=9400$; (c) $N=9390$. }
         \centering
        \begin{tabular}{|c|c|}
        \hline
         (a)  \includegraphics[width=0.4\textwidth]{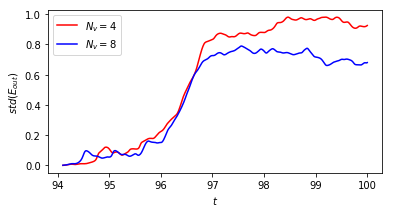}  &
         (b) \includegraphics[width=0.4\textwidth]{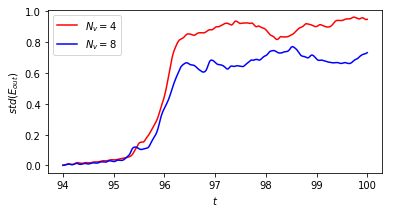}  \\ 
            \hline
        \end{tabular}
        \label{std_eg2}
                \begin{tabular}{|c|}
        \hline
         (c)  \includegraphics[width=0.4\textwidth]{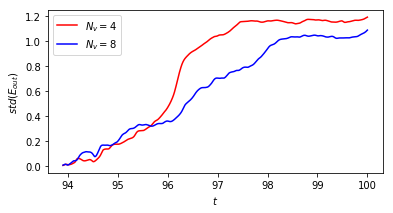} \\ 
            \hline
        \end{tabular}
                \label{std_eg2}
    \end{table*}

Figure (1b) and Figure (2b) in Table \ref{res_eg2} show that our method is capable of quite confidently predicting the rare event and its transition path at least 216 numerical time steps in advance, or at least about 30 autocorrelation time of the fast driving signal prior to first cross-over of $x$ is at $t = \tau_1$. Even though the prediction results in Figure (1a) and Figure (2a) are inferior to those with the larger $N_v$, they are not too far behind since the cross-over to the origin is successfully predicted there.

However, the figures (3a)-(3b) in Table \ref{res_eg2} show how the inability of the DESN to make accurate long-term predictions again cripples the performance of the prediction method, making prediction of the rare event from an earlier time a very challenging task. In particular, neither models can confidently predict the cross-over of $x$ to the origin during the prediction interval, although the result obtained by the model with $N_v=4$ comes closer to achieving that. We emphasize that these results are used primarily to demonstrate our method. It may be possible to improve these results by using a different training setting (with a higher number of layers and more carefully optimizing values of the hyperparameters), enabling successful longer-term prediction and therefore prediction of a rare event from an earlier time. 

These figures again depict the difference in the prediction results obtained with different values of $N_v$ used (see the figures in Table \ref{error_eg2}). Just like the case in Example 1, the qualitative behaviors of both the mean value and standard deviation of the error of the predicted values are essentially the same for different $N_v$. However, the standard deviation plateau is higher in all the cases where $N_v$ is lower (see Table \ref{std_eg2}), in contrast to the finding in the case of Example 1. This indicates that larger values of $N_v$ should be considered here to obtain a prediction model with a lower variance.

We find that it is very difficult to obtain a comparable quality of predictions if we work with a shallow ESN, instead of the three-layered variant that we have used here. This finding may  be explained by the following. A closer look at the model for the driving signal $z(t)$  reveals that there are actually two widely separated time scales in the system generating it, instead of only one time scale as in the case of Example 1. This difference in the multiscale behavior of the system generating the driving signal explains why a shallow ESN works well for Example 1 but not for Example 2. A deep version of the ESN is needed to handle the multiscale nature of the data for $z$ in Example 2. This supports the intuition that increasing the depth of the ESN can lead to a better multiple time scale representation of the temporal information.  This motivates our consideration of a deep version of the ESN for our method in Section \ref{sect_alg}.

The figures (0a)-(0b) in Table \ref{res_eg2} confirm that our method is far superior to the baseline method of applying a DESN to the data for $x$. Indeed, the ability of our method to predict, at least 216 time steps (at least 2 periods) in advance, has significant importance for many systems exhibiting stochastic resonance. It is possible to achieve prediction results of comparable quality using a shorter but sufficiently long trajectory starting at a later time $t_{init}>0$ and coming sufficiently close to the rare event (see the relevant discussion on this in Subsection \ref{eg1}). 

\begin{widetext}
\subsection{Example 3: A tri-stable system, subject to periodic forcing and driven by a multiscale Lorenz-96 system}

\noindent {\bf Data generation.} The data for $x$ is generated by the following slow-fast system:
\end{widetext}

\begin{widetext}
\begin{align}
\dot{x}(t) &= x(t)[1-x(t)][1+x(t)][x(t)-2][x(t)+2] + B \cos(\Omega t) + \sigma \xi_1(t), \label{Ex3_1} \\ 
\dot{\xi}_1 &= \frac{1}{\epsilon}((\xi_2-\xi_{K-1})\xi_K) - \xi_1 + F, \\ 
\dot{\xi}_2 &= \frac{1}{\epsilon^2}((\xi_3-\xi_{K})\xi_1 - \xi_2 + F), \\
\dot{\xi}_K &= \frac{1}{\epsilon^2}((\xi_1-\xi_{K-2})\xi_{K-1} - \xi_K + F), \\
\dot{\xi}_{n} &= \frac{1}{\epsilon^2}((\xi_{n+1}-\xi_{n-2})\xi_{n-1} - \xi_n + F), \ \ \text{ for } n=3, \dots, K-1. \label{Ex3_5}
\end{align}
\end{widetext}

\begin{figure}[!h]
\caption{{\it Time series data for $x$ in Example 3 up to time $t=100$}, of which only a segment prior to $t=82.33  < \tau_1 = 82.39$ (i.e., at most $N+1=8234$ data points) will be available for training. Note that the sample path of $x$ first crosses one at $t = \tau_1$ (indicated by the yellow dashed line), where a critical transition occurs.}
\centering
\includegraphics[width=0.48\textwidth]{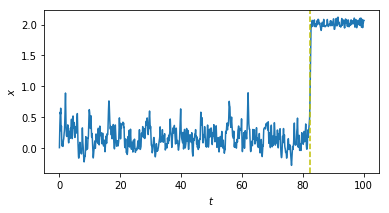}
\label{eg3_whole}
\end{figure}

In \eqref{Ex3_1}-\eqref{Ex3_5}, $x$ is the state of the system of interest whose evolution is driven by a fast $K$-dimensional Lorenz-96 system whose state is denoted by the vector $(\xi_1, \xi_2, \dots, \xi_K)$. We work with $K=36$ and $F=8$, which are parameter values leading to chaotic dynamics \cite{lorenz1996predictability}. For data generation, we use a uniform time step of $\Delta t = 0.01$ to integrate the above system with $x(0) = 0$, $\xi_k(0) = F = 8$ for all $k$ except for $k=20$, with $\xi_{20}(0) = F+0.01$, $B=0.1$, $\Omega = 12 \pi$, $\sigma = 0.4$ and $\epsilon = 0.5$, up to time $t=100$. The autocorrelation time  of samples of the driving signal is estimated to be $0.11$.  The time step has again been chosen to be small enough to sample the scale on which the fast signal takes place.  The time series generated for $x$ is plotted in Figure \ref{eg3_whole}. We assume that we only have access to a segment of the {\it single} time series plotted in Figure \ref{eg3_whole} up to time $t<\tau_1$.\\

\noindent {\bf Generating system: dynamics and applications.} The system  described by \eqref{Ex3_1}-\eqref{Ex3_5} is similar to the one in Example 2 except that it is driven by a different periodic forcing and  the fast driving signal comes from a single component of a variant (due to the scaling introduced by $\epsilon$) of the Lorenz-96 system, a much higher dimensional chaotic system compared to the ones considered in the previous  examples.   We remark that, due to the scaling introduced, the limiting slow dynamics are deterministic (averaging) rather than stochastic (homogenization). The critical transitions in this multiscale system \eqref{Ex3_1}-\eqref{Ex3_5} are induced by finite $\epsilon$-effects which are fluctuations around some constant mean driver. 

The Lorenz-96  model (with $\epsilon = 1$) above is the first model introduced by Lorenz in  \cite{lorenz1996predictability}. It is a model extensively used in data assimilation and parameter estimation (parametrization) research, as well as in testing machine learning algorithms for parameter learning (sub-grid parametrization) \cite{vlachas2020backpropagation,gmd-2019-136}.  It was originally used by Lorenz as a one-dimensional atmospheric model. The variables in the model represents values of some atmospheric quantity in $K$ sectors of a latitude circle, giving a periodic system of $K$ ODEs. The basic physics of the atmosphere is captured in the right hand side of the ODEs, which contains  advection terms, damping terms, and  external forcing.   One can therefore view $x$ as a sea-surface temperature anomaly, influenced by the atmospheric quantity in one of the sectors of the latitude circle in the presence of a seasonal forcing. From the data, we find that the first crossing time of the sample path to one is $\tau_1 := \inf\{t \in [0,100] : x(t) > 1\} = 82.39$.  \\

\noindent {\bf Results and discussion.} Figure (1a) in Table \ref{res_eg3} shows that our method is capable of confidently predicting the rare event and its transition path at least 6 time steps in advance, prior to first crossing of the slow process to one at $\tau_1= 82.39$. This is equivalent to about 0.55 autocorrelation times of the fast driving signal. The target future trajectory lies mostly  within the $90$ percent confidence interval of the predicted trajectory. This result is  far better than the direct method of applying an ESN to the data for $x$ (see the figures in (0a)-(0b)), which we take as the baseline results.


\begin{table*}
 \caption{{\it Prediction results using the data set in Example 3 under various scenarios.} \\
$N$ is the number of data points in the training set, $N_v$ is the number of data points in the validation set (see Appendix). 
In each figure: the target trajectory for $x$ (in red), $N_{eff}$ predicted trajectories from $N_{eff}$ independently trained and selected models, and the averaged predicted trajectory (in thick blue) up to several time steps into the future. The shaded light blue region represents the  $90\%$ confidence interval of the predicted values.}
        \centering
        \begin{tabular}{|c|c|c|}
        \hline
           \toprule
            Scenarios & $N_v = 5$ & $N_v = 8$  \\
            \hline
            \midrule
            Baseline  & (0a) \includegraphics[width=0.4\textwidth]{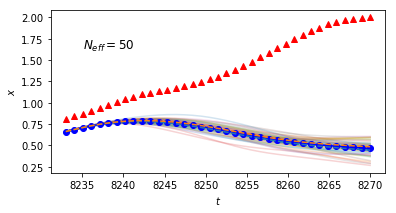} & (0b) \includegraphics[width=0.4\textwidth]{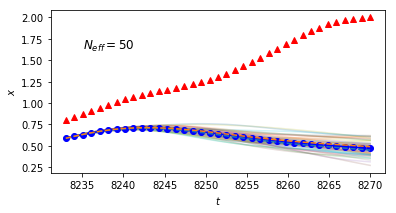} \\ \hline
            $N=8233$ & (1a) \includegraphics[width=0.4\textwidth]{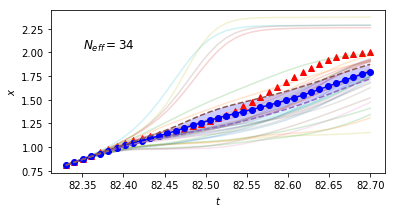} & (1b) \includegraphics[width=0.4\textwidth]{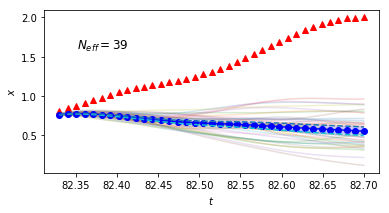} \\ \hline
            $N=8231$ &  (2a) \includegraphics[width=0.4\textwidth]{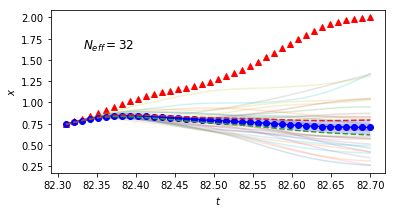} & (2b) \includegraphics[width=0.4\textwidth]{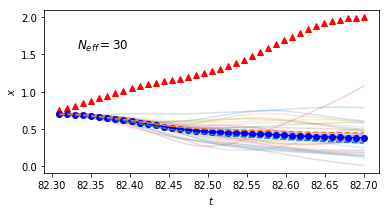} \\ \hline
            \bottomrule
            \hline
        \end{tabular}
        \label{res_eg3}
    \end{table*}

\begin{table*}
 \caption{{\it   Error of the predicted results and its mean value (in thick blue) for Example 3. }}
         \centering
        \begin{tabular}{|c|c|c|}
        \hline
           \toprule
            Scenarios & $N_v = 5$ & $N_v = 8$  \\
            \hline
            \midrule
            $N=8233$ &  (1a) \includegraphics[width=0.4\textwidth]{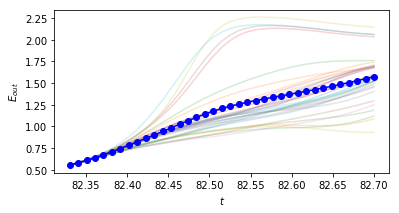} &  (1b) \includegraphics[width=0.4\textwidth]{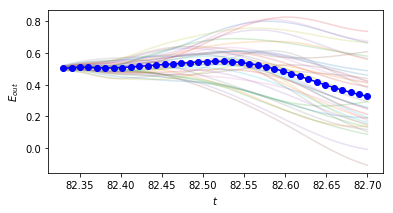} \\ \hline
            $N=8231$ &  (2a) \includegraphics[width=0.4\textwidth]{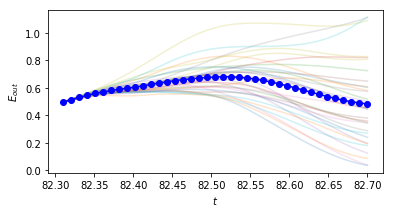} & (2b) \includegraphics[width=0.4\textwidth]{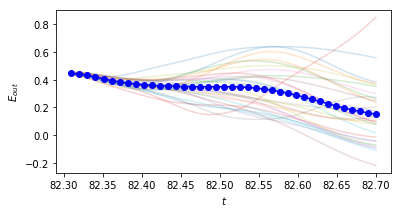} \\ 
            \bottomrule
            \hline
        \end{tabular}
        \label{error_eg3}
    \end{table*}

\begin{table*}
 \caption{{\it  Standard deviation of the errors for Example 3. } 
(a) $N=8233$; (b) $N=8231$. }
         \centering
        \begin{tabular}{|c|c|}
        \hline
         (a)  \includegraphics[width=0.4\textwidth]{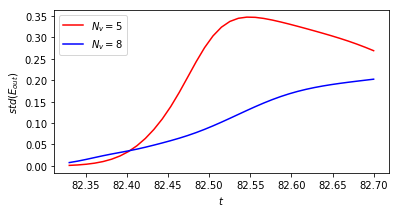}  &
         (b) \includegraphics[width=0.4\textwidth]{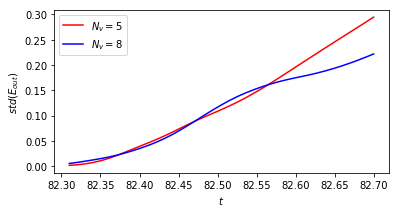}  \\ 
            \hline
        \end{tabular}
        \label{std_eg3}
    \end{table*}

Figure (1b) and Figure (2a)-(2b) in Table \ref{res_eg3} show that the prediction models are unable to anticipate the rare transition event in the prediction interval. Therefore, in comparison to the results in earlier examples, here the predictive performance of our models are far inferior. However, we still manage to produce an accurate prediction model as shown in Figure (1a).  This is despite the fact that we are working with a set of data extracted only from one of the components  of the Lorenz-96 system and moreover these data are noise corrupted -- a highly challenging task even in the case when the data is noise-free (see  \cite{vlachas2020backpropagation}). We have also attempted to use deeper versions of network for training but, unlike the case of Example 2, we are unable to obtain better predictions than those obtained by shallow networks.

Figure (1a)-(2b) in Table \ref{res_eg3}  depict the difference in the prediction results obtained with different values of $N_v$ used, as discussed before. Unlike the earlier examples, the qualitative behavior of the mean value and standard deviation of the error of the predicted results are not the same for different $N_v$ in the case when $N=8233$ (see the figures in Table \ref{error_eg3} and Table \ref{std_eg3}). This indicates high sensitivity of the prediction models to the choice of $N_v$ when dealing with highly complex multiscale data. We suspect that such sensitivity is due not only to the sensitivity of the echo state network to the choice of parameters, but also due to the noise corrupted nature of the extracted data. It is interesting to observe that the successful prediction model described here produces predictions whose standard deviation of the error shows a peak on the prediction interval, whereas those for the unsuccessful ones have strictly increasing standard deviation on the interval.

\section{Conclusions}
\label{sect_concl}

We have presented a data-driven method for tackling the task of predicting noise-induced critical transition events in a large class of multiscale nonlinear dynamical systems. The method was demonstrated on three examples of different complexity, each providing a model for many important physical and biological systems. For Example 1 and Example 2, we obtain excellent predictions that would not be possible by using a direct method that does not take into account the multiscale nature of the problem. In particular, our method successfully predicts a rare transition event up to several numerical time steps  in advance. Each run of the ESN/DESN incurs relatively low computational cost, thanks to the use of a reservoir computing based training technique, rather than a gradient descent based method. This low computational cost allows us to leverage the power of ensemble learning for the predictions. 

These results demonstrate the promise of our  reservoir computing based method in predicting rare events occurring in a wide range of  dynamical systems, a  problem that is of substantial interest in science and engineering. We expect the accuracy of these results to improve by carefully optimizing  the hyperparameters, using a more refined training method and  other more sophisticated machine learning algorithms, at the expense of a higher computational cost. Importantly, rather than demonstrating only the successful results, we also highlight the limitation of the approach when dealing with highly complex multiscale data sets, such as that shown in Example 3. We also empirically show and discuss the sensitivity of the method to parameters and training data. Apart from achieving successful predictions, understanding and mitigating these issues are equally important. Therefore, our findings open up a range of interesting problems for future work.

We now discuss a few potential future directions. Thus far we have applied the method to ``toy examples'', where there are few widely separated time scales. In many systems of interest there may be many widely separated time scales and the competition between them may be crucial in triggering a critical transition  event. The driving noise may also be multiplicative in nature. Therefore, it is important to extend the present work to these systems. In many realistic situations, the available time series data may be multivariate, rather than univariate, and additionally there may be missing and/or uneven data. It would then be important to extend our method to treat these situations.

It is also of practical interest to apply the method presented here to study more non-trivial yet physically relevant data sets, such as those generated by a chaotic version of the model in \cite{eisenman2009nonlinear} and/or real world data from climate science \cite{2019arXiv190605433R,Scher18}. On the other hand, because the method is based on the use of (a deep version of) echo state network, a firm theoretical understanding of the underpinnings behind such a network, in particular its initialization (e.g., the role of randomness \cite{scardapane2017randomness}) and the generalization error, will  shed light on the nature of the prediction results. Therefore, it is important to carry out a systematic theoretical study to understand how the network works. This is in fact an important problem in the field of reservoir computing and vigorous efforts have been made  in recent years to tackle this task (see, for instance, \cite{gonon2019risk,hart2020embedding,gonon2020approximation,cuchiero2020discrete,verzelli2020input,verzelli2020learn}).  \\

\begin{widetext}
\section*{ Data Availability}

The complete codes (in Python) that reproduce all the results obtained in this paper are openly available in GitHub at:\\ 
 \url{https://github.com/shoelim/predicting-rare-critical-transitions-in-multiscale-systems}

\begin{acknowledgments}
The authors acknowledge Swedish Research Council grant no. 638-2013-9243. S.H. Lim is grateful to Stefano Bo for critical reading of the manuscript and many insightful discussions. 
\end{acknowledgments}
\end{widetext}

\section*{Appendix: Details on Training, Model Selection and Testing}

We adopt the following approach for training, validation (model selection) and testing for the three examples studied in the paper.

From a given time series for the slow variable $x$ (consisting of at least $N+1+M$ data points), we assign a training set, a validation set (with a total of $N+1$ data points) and a test set (with $M$ data points). The training set and validation set are the only accessible (in-sample) data and we are {\it not allowed} to use the (out-of-sample) data in the test set, which records the occurrence of a rare event. From these accessible data, we first extract the driving signal according  to (S1) in Algorithm \ref{thealg}. There will be $N$ data points for the signal. We set aside the last $N_v$ data points of the extracted signal on the time interval $[(N-N_v)\Delta t, (N-1) \Delta t]$ to build a validation set, which will be used for hyperparameter tuning and model selection.  

For each realization of the ESN/DESN, our  model selection procedure is as follows. We employ the classical static validation scheme (see \cite{cv_esn19} for details) and perform a grid search over a pre-chosen hyperparameter space, looking for hyperparameter values that minimize the root mean squared error (RMSE) on the validation set.  To choose this space, we focus on optimizing over the number of reservoir elements ($n_{x_i}$) in each layer, while fixing the desired spectral radius ($\rho_{des}^{(i)}$), noise intensity ($\nu$), and sparsity parameter ($r_i$) to reasonable default values. This choice is informed by the practice that optimizing over $n_{x}$ (the memory capacity) should be prioritized over other hyperparameters \cite{lukovsevivcius2012practical}. The values of all these hyperparameters are not fully optimized, i.e., an exhaustive search over the hyperparameters has not been performed. However, good average performance is typically not found in a very narrow parameter range, thus a very detailed fine-tuning of parameters does not give a significant improvement and is not necessary \cite{lukovsevivcius2012practical}.  An alternative model selection procedure is the more expensive automated Bayesian optimization approach proposed in \cite{yperman2016bayesian}. 

The optimal ESN/DESN models selected by the above procedure will depend on the choice of $N_v$. The choice of $N_v$ should be guided by the size and characteristic of the accessible data: $N_v$ needs to be sufficiently large so that the validation error is a reliable estimate of out-of-sample error but should be small enough, keeping in mind the size of accessible data and the short-term forecasting capability of ESN/DESN. Different choices of $N_v$ will therefore give rise to different prediction models.  Once an optimal model is chosen, we run the trained network  to obtain predicted values for the driving signal on the time interval $[(N-N_v)\Delta t, (N+M)\Delta t]$ (see (S2) in Algorithm \ref{thealg}) and proceed to (S3)  in Algorithm \ref{thealg}.  For numerical integration in (S3), we use a Runge-Kutta method with a uniform step size of $\Delta t = 0.01$.

Our prediction and testing procedure is as follows. We perform an ensemble-like prediction by repeating the above training and model selection procedure $N_{ens}$ times, obtaining $N_{ens}$ independently trained and optimized prediction models, whose networks are initialized using different random seeds. The idea of this ensemble algorithm is that independent training and model selection here produce different learners (committee of networks), learning on a diverse set of features. The predictions of several base models built with the same algorithm are then aggregated in order to improve robustness over a single model; there may be predictions that are completely off but these will partially cancel each other out. This increases stability, lowers the variance and optimizes the overall predictive performance. A higher (lower) $N_{ens}$ gives a more (less) confident estimator. The choice of $N_{ens}$ should not be too large to minimize computational expense in training models and to avoid diminishing returns in performance from adding more ensemble members. We choose $N_{ens} = 50$ in all examples here. We have repeated the experiments with a larger $N_{ens}$ of 100 but find only  slight improvement in performance. 

Since some members of the ensemble will  make better predictions than others, we expect to reduce the test error further if we assign greater weight to some members than to others. For this reason, we are taking weighted averaged values as the final predicted values as follows. Let $N_{eff}$ be the number of predicted trajectories that lie entirely inside a pre-chosen confidence interval (which should also cover the trajectory obtained from the validation set) around the mean predicted trajectory on the time interval defining the validation set. The weight ascribed to each predicted trajectory is set to $1/N_{eff}$ (getting a vote) if it falls entirely inside the confidence interval, and  zero (getting no vote) otherwise. This filters out the outliers, leaving us with $N_{eff} \leq N_{ens}$ selected  models that are expected to have reasonably good short-term forecasting capability, at least on the validation set.  To compute the ensemble average in our experiments, we choose the confidence interval to be  within 8 standard deviations of the mean predicted trajectory on the validation time interval (if none of the predicted trajectories falls inside this region, we increase the number of standard deviations used by one iteratively until at least one falls inside the region). 

\begin{figure*}
\caption{{\it A schematic summarizing the workflow of our data-driven method for rare event prediction.} In the schematic below, (S1)-(S3) denote the procedures described in Algorithm \ref{thealg}, $M$ denotes a (realization of) ESN/DESN model described in Section \ref{sect_esn} during the training phase, $P$ denotes a prediction model obtained by running a ESN/DESN model autonomously during the prediction phase, $M^*$ denotes an optimal ESN/DESN model produced by our model selection procedure, $N$ is the number of data points in the training set, $N_v$ is the number of data points in the validation  set, $N_{ens}$ is the number of learners in an ensemble,  $N_{eff}$ is the number of predicted paths that lie entirely inside a pre-chosen confidence interval around the mean predicted path on $[t_{N-N_v}, t_{N-1}]$, and the $\hat{x}_i$ denote the predicted values obtained by the chosen learners in the ensemble.  In the plot (not drawn to scale) for the final prediction model, the target path $(x)$ is in red and the predicted paths ($\hat{x}_i$) are in blue. We take the averaged predicted paths (over the $N_{eff}$ blue paths) as the final prediction model. }
\centering
\includegraphics[scale=0.75]{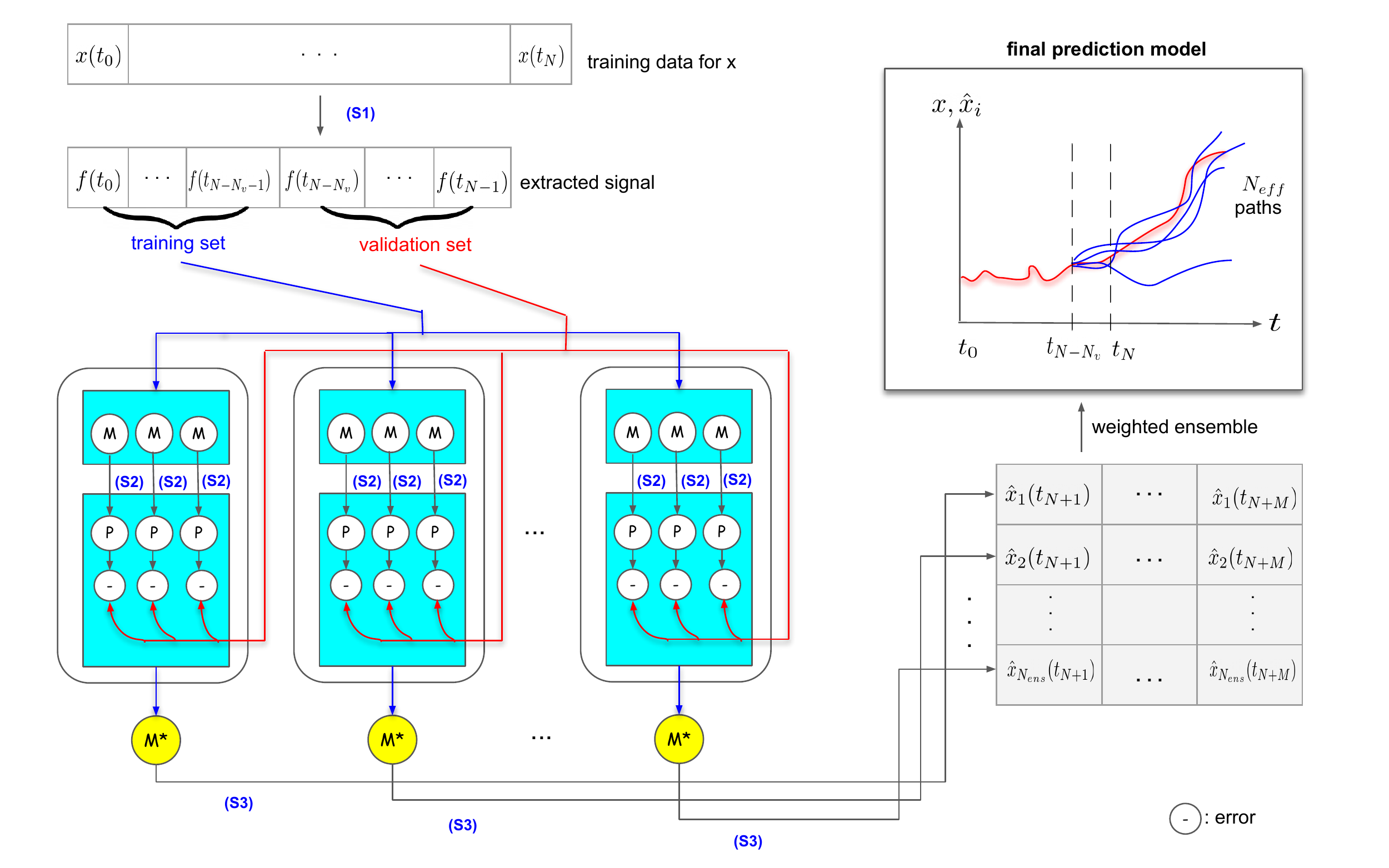}
\label{ms}
\end{figure*}

To evaluate the quality of  the predictions of the $N_{eff}$ models, we study the statistics (the mean and standard deviation) of the difference  between the predicted values and the target (real) values. Note that alternative measures for evaluating the predictive performance could also be considered. Our choice of measure here, in contrast with coarse-grained metrics such as the percentage of predicted trajectories that display a jump at a time close to the real one, reflects our focus on inferring the characteristics of the whole transition path.  Figure \ref{ms} shows a schematic of our method. 

The following are implementation details specific to each example.

\begin{itemize}
\item {\bf Example 1.} We apply Algorithm \ref{thealg}-\ref{alg_desn} using a shallow ESN (i.e., $n_L=1$ and $L=0$) with $n_u = n_y=1$, the bias value of $b=1$, and $i_{transient} = 0$. The size of the validation set is chosen to be $N_v = 10$ and $N_v = 20$. For the grid search, the range of values that we use for $n_{x_0}$ are from $600$ to $740$, while  $\rho^{(0)}_{des} = 0.7$, $r_0 = 0.1$, and $\nu = 0.001$.

\item {\bf Example 2.}
We apply Algorithm \ref{thealg}-\ref{alg_desn} using  a three-layered DESN (i.e., $n_L=3$, $L=2$) with $n_u = n_y=1$,   the bias value of $b=1$, and $i_{transient} = 0$.  The size of the validation set is chosen to be $N_v = 4$ and $N_v = 8$. For the grid search, the range of values that we use for $n_{x_i}$ (for $i=0,1,2$) are from $100$ to $300$, while  $\rho_{des}^{(0)} = 0.6$, $\rho_{des}^{(1)} = 0.7$, $\rho_{des}^{(2)} = 0.8$, $r_0 = r_1 = r_2 = 0.05$, and $\nu = 0.003$.

\item {\bf Example 3.} We apply Algorithm \ref{thealg}-\ref{alg_desn} using a shallow ESN (i.e., $n_L=1$ and $L=0$) with $n_u = n_y=1$ and the bias $b=1$. The first 200 data points are not used (so $t_0 = 2$) and the first $i_{transient} = 100$ hidden states of the ESN are discarded during training. The size of the validation set is chosen to be $N_v = 5$ and $N_v = 8$. For the grid search, the range of values that we use for $n_{x_0}$ are from $500$ to $700$, while  $\rho^{(0)}_{des} = 0.95$, $r_0 = 0.1$, and $\nu = 0.003$.

\end{itemize}

We invite interested reader to experiment with the choice of hyperparameters using the codes provided at the website in {\bf Data Availability}.




\section*{References}
\bibliographystyle{siam}


\end{document}